\documentclass[aps,pra,twocolumn,a4paper,amsmath,amssymb,%
showpacs,floatfix]{revtex4}
\usepackage{graphicx}
\usepackage{epstopdf}

\newcommand{\Op}[1]{\boldsymbol{\mathsf{\hat{#1}}}}

\newcommand{\Fkt}[1]{\,\mathsf {#1}}

\newcommand{\eps}{\varepsilon}
    
\begin{document}
\hyphenation{Fesh-bach}
\title{Short-pulse photoassociation in rubidium below the D$_1$ line}
\date{\today}
\author{Christiane P. Koch}
\email{ckoch@fh.huji.ac.il}
\affiliation{Department of Physical Chemistry and
  The Fritz Haber Research Center, 
  The Hebrew University, Jerusalem 91904, Israel}
\author{Ronnie Kosloff}
\affiliation{Department of Physical Chemistry and
  The Fritz Haber Research Center, 
  The Hebrew University, Jerusalem 91904, Israel}
\author{Fran\c{c}oise Masnou-Seeuws}
\affiliation{Laboratoire Aim\'e Cotton, CNRS, B\^{a}t. 505, Campus d'Orsay,
91405 Orsay Cedex, France}

%\pacs{33.80.-b,32.80.Qk,34.50.Rk,33.90.+h}
\pacs{33.80.Ps,32.80.Qk,34.50.Rk}

\begin{abstract}
Photoassociation of two ultracold rubidium atoms
and the subsequent  formation of stable molecules in the singlet ground
and lowest triplet states
is investigated theoretically. The method employs  laser pulses  inducing
transitions via excited states correlated to the $5S+5P_{1/2}$ asymptote.
Weakly bound molecules in the singlet ground
or lowest triplet state can be created by a single
pulse while the formation of more deeply bound molecules 
requires a two-color pump-dump scenario. More 
deeply bound molecules in the singlet ground or lowest triplet state
can be produced only if
efficient mechanisms for both pump and dump steps exist.
While long-range $1/R^3$-potentials allow for efficient photoassociation,
stabilization is facilitated by the resonant spin-orbit coupling
of the $0_u^+$ states. Molecules in the singlet ground state bound
by a few wavenumbers can thus be formed. This provides a promising first
step toward ground state molecules which are ultracold in both translational and
vibrational degrees of freedom.
\end{abstract}

\maketitle

%--------------------------------------------------------------------------------%
\section{Introduction}
\label{sec:intro}

The formation of ultracold molecules along with the creation of molecular
Bose-Einstein condensates (BEC)~\cite{JochimSci03,GreinerNat03,ZwierleinPRL03}
opens the way to a new field of research~\cite{Quovadis}.  %of  ultracold chemistry
Methods to directly cool the translational degrees of freedom of molecules have
advanced significantly over the last few years.
However, the lowest temperatures to date have  been achieved
by cooling atoms and then assembling the atoms into molecules
by applying an external field. To this end,
magnetic as well as optical fields have been employed.
The formation of weakly bound alkali dimer molecules with magnetic fields
using so-called Feshbach resonances has been particularly
successful, see e.g.~\cite{StreckerPRL03,XuPRL03,HerbigScience03}.
Optical techniques, however,  present a very general technique due to the abundence
and diversity of optical transitions. Furthermore, laser fields
can be manipulated offering various possibilities for control.
In particular, molecules in their vibronic ground state can be
created by employing a combination of light fields of different colors~\cite{SagePRL05}.
The major drawback of optical techniques is the fairly short lifetime
of the electronic states involved, on the order of a few ten's of nanoseconds
for the alkali dimer systems. The coherence of the control process may then be lost
because of spontaneous emission~\cite{RomPRL04,McKenziePRL02}. This flaw
which is particularly severe in view of creating a stable molecular BEC, 
can be overcome by employing laser pulses with a duration shorter than the
excited state lifetimes.

The use of short laser pulses offers several advantages 
in the manipulation of ultracold systems. Obviously the
shorter timescale can overcome the losses induced by spontaneous emission.
In addition, a sequence of shaped pulses can control the photoassociation mechanism.
Two time-delayed pulses can be employed to realize pump-probe experiments:
The first or pump pulse creates a wavepacket whose dynamics is
monitored by a second  probe pulse, varying the time
delay~\cite{GensemerPRL98,FatemiPRA01}.
The shape of the pulse can be manipulated, for example
by introducing a controlled frequency chirp which enforces adiabatic following
conditions~\cite{JiriPRA00,ElianePRA04,ElianeEPJD04,WrightPRL05}. More elaborate coherent control
which shapes amplitude and phase can also be employed.
These pulses can be found using feedback learning loops~\cite{BrixnerCPC03}. 
The underlying idea is  to employ information
on quantum interferences to construct a pulse which is able 
to steer the process into the desired direction~\cite{RabitzScience00}.
However, the know-how of closed-loop, coherent control
experiments~\cite{BrixnerCPC03} cannot be transferred to experiments
on ultracold systems in a one-to-one fashion. This is because the dynamical timescales of cold
systems are simply much larger than the ultrashort  timescale of femtoseconds.
The adaptation of pulse shaping techniques known from coherent control
to ultracold systems is  a subject of current
investigation~\cite{Salzmann,Brown}. 
In addition to controlling nuclear dynamics, 
a femtosecond laser can be employed for high-resolution spectroscopy
in ultracold systems by utilizing it as a frequency comb~\cite{UdemPRL99}.
These two limits have recently been combined in a spectroscopic
study with coherent pulse accumulation~\cite{YeScience04}. The appeal
of such an approach lies in the fact that  with a femtosecond frequency comb
it is possible to continuously switch between the
two limits, selecting the desired time and frequency resolution.

Photoassociation (PA) has originally been developed
with continuous-wave (CW) lasers, allowing for
high-precision spectroscopy~\cite{Weiner98,FrancoiseReview}.
PA with short laser pulses does not yield spectroscopic
information and requires theoretical modelling. In general, PA
is defined as  the formation of electronically excited
molecules from two colliding atoms by interaction with laser light. In a 
subsequent stabilization step,
molecules in the singlet ground state or in the lowest triplet state
are created by spontaneous or stimulated emission.
Photoassociation with short laser pulses on cold systems 
has sparked interest from both 
experiment~\cite{SagePRL05,WrightPRL05,Salzmann,Brown}
and theory~\cite{MachholmPRA94,VardiJCP97,JiriPRA00,ElianePRA04,ElianeEPJD04,My05}.
Specifically, a pump-dump scheme has been suggested with
the idea to create a wavepacket on the excited state
and make use of its time-dependence for efficient
stabilization~\cite{ElianePRA04,ElianeEPJD04,My05}.
In order to obtain an overall high efficiency, both
excitation (photoassociation) and deexcitation (stabilization)
steps must be efficient. 
PA can be understood as a vertical transition at
the Franck-Condon distance $R_C$
depending on the frequency $\omega_P$ or detuning $\Delta_P$.
It works best at long range~\cite{Weiner98,FrancoiseReview}.
Since excitation of the atomic resonance needs to be avoided, this
requires a rather narrow bandwidth, i.e.
pulses in the picosecond to nanosecond regime.
The efficiency of the stabilization step depends on the topology of the
potentials and the time-delay between pump and dump pulses~\cite{My05}.
In PA experiments with CW lasers, the stabilization step is determined
only by the potentials and the transition dipole moment,
i.e. by the Franck-Condon factors of single
excited state vibrational levels. In a time-dependent process the
excited state wavepacket is made up of a superposition of several
vibrational levels and moves under the influence of the excited state potential.
This movement can also be viewed as interferences in time of the
superposition coefficients. Stabilization is most effective if the dump
pulse interacts with the molecular wavepacket when it is located 
at short internuclear distances. For cesium it was found that
up to 20\% of the excited state wavepacket can be
transferred to molecules in the lowest triplet state~\cite{My05}.

The present study employs the concepts developed in
Refs.~\cite{JiriPRA00,ElianePRA04,ElianeEPJD04,WrightPRL05} for
PA with chirped pulses and  of Ref.~\cite{My05} for the pump-dump scheme.
In previous work on cesium, only 
excitation of the $0_g^-(P_{3/2})$ excited state
and molecule formation in the lowest triplet state $a^3\Sigma_u^+$ 
were considered~\cite{My05}. 
In order to  analyze short pulse PA from the
point of view of a prospective experiment, the present study goes beyond the two-state
description~\cite{JiriPRA00,ElianePRA04,ElianeEPJD04,WrightPRL05,My05}.
Formation of molecules in both the singlet ground state $X^1\Sigma_g^+$ 
and the lowest triplet state $a^3\Sigma_u^+$ can then be tested.
Rubidium has been chosen since experimental effort is under way for this
species~\cite{Salzmann,Brown}.
A comprehensive model of the interaction of two atoms with a laser field
is developed and the role
of all laser parameters is analyzed. In particular, due to the bandwidth of the pulse,
a number of vibrational levels of the electronically excited state are excited. These 
levels may belong to different electronic states. For this reason  
all potentials which allow for transitions within the bandwidth of the pulse are
included in the model. 
For the sake of brevity, the focus is on the potentials 
correlated to the $5S+5P_{1/2}$ atomic asymptote, i.e. to the $D_1$ line.
The corresponding model will be introduced in Sec.~\ref{sec:model}.
Considering transitions into $5S+5P_{1/2}$ potentials implies that the
central frequency of the laser is red-detuned from
the $D_1$ line. The control knobs which can be varied
in the experiment are the parameters of the laser pulse, i.e. its
central frequency, spectral bandwidth, intensity and possibly frequency chirp.
The variation of these parameters will guide the discussion of the mechanisms of
excitation and deexcitation in Secs.~\ref{sec:excitation} and
\ref{sec:deexcitation}, respectively. The separation  of pump and dump
mechanisms is not only a convenience of the theoretical discussion.
It is motivated by the fact that
the two steps correspond to two different detection schemes in an experiment:
The excitation of two colliding atoms into
long-range, electronically excited molecules would be measured as trap loss
of the atomic cloud while the formation of molecules
in the singlet ground and lowest triplet states would be monitored
by R(esonantly) E(nhanced) M(ulti)P(hoton) I(onization) spectroscopy.
Accurate information from \textit{ab initio} calculations
and spectroscopy for the molecular potentials and transition dipole moments
is employed in the current study.
Unfortunately, no such data is  available as yet for the description
of the spin-orbit coupling. Therefore three different
model curves are employed, and the sensitivity of the proposed scheme with respect
to the spin-orbit coupling is discussed in Sec.~\ref{sec:SO}.
Finally, conclusions are drawn in Sec.~\ref{sec:concl}. In particular, 
the experimental feasibility of
creating ultracold stable molecules with short pulse PA is discussed.

%-------------------------------------------------------------------------------------
\section{Model}
\label{sec:model}

Two colliding rubidium atoms in their ground state ($5S+5S$) 
interacting with a laser field are considered. 
The laser excites the two atoms into an electronically excited state
($5S+5P$) which may support long-range molecular bound levels. Four $\Sigma^+$ and four
$\Pi$ (Hund's case (a)) states (singlet/triplet and gerade/ungerade) are correlated to the
$5S+5P$ asymptote with two of them repulsive and two attrative.
When including the spin-orbit interaction (see e.g.~\cite{WangJCP96}),
the respective $0$, $1$ and $2$ (Hund's case (c))
potentials are obtained. Five of these are correlated
to the $5S+5P_{1/2}$ asymptote, i.e. the $D_1$ line. Out of these five,
only the attractive potentials $0_u^+(P_{1/2})$, $1_g(P_{1/2})$
and $0_g^-(P_{1/2})$ which support bound molecular levels are considered.
For the $0_u^+$ states, the spin-orbit coupling has resonant
character~\cite{AmiotPRL99,SlavaJCP99,DulieuJOSA03} and leads to an avoided crossing between
$0_u^+(P_{1/2})$ and $0_u^+(P_{3/2})$ at short internuclear
distance. The spin-orbit coupling therefore needs to be included
explicitly: The Hund's case (c) potentials are obtained
by diagonalizing the Hamiltonian containing the Hund's case (a) potentials
on the diagonal and the spin-orbit coupling on the off-diagonal~\cite{WangJCP96}.
The corresponding unitary transformation leads to off-diagonal elements of the kinetic
energy which usually are neglected. For resonant coupling, however,
these 'non-adiabatic couplings' have to be taken into account: 
In the case of the two
$0_u^+(P_{3/2})$  and $0_u^+(P_{1/2})$ curves, it
couples the two vibrational series and leads to
well-known perturbations in the
spectra~\cite{AmiotPRL99}.

Due to conservation of the  gerade/ungerade symmetry in
homonuclear dimer molecules, optical transitions
are allowed to $0_u^+$ from the singlet $X^1\Sigma^+_g$ ground state
and to $1_g$ and $0_g^-$ from the $a^3\Sigma^+_u$ lowest triplet state.
For the interaction with the field, the
dipole and rotating wave approximations are assumed.
The $0_u^+(P_{1/2})$, $1_g(P_{1/2})$ and $0_g^-(P_{1/2})$ states on one hand
and the  $X^1\Sigma^+_g$ and  $a^3\Sigma^+_u$ 
states on the other hand are coupled by hyperfine interaction.
However, since the
timescale associated with the hyperfine interaction is much larger
than the timescales of femtosecond and picosecond pulses,
the coupling between the states can be neglected, and
three separate Hamiltonians
$\Op{H}_1$,  $\Op{H}_2$ , and  $\Op{H}_3$ will be considered.

The Hamiltonian $\Op{H}_1$ which describes transitions from the singlet ground state
into the $0_u^+$ excited states is given by
\begin{widetext}
  \begin{equation}
    \label{eq:Ham1}
    \Op{H}_1 =
    \begin{pmatrix}
      \Op{T} + V_{X^1\Sigma_g^+}(\Op{R}) & \mu_\pi(\Op{R}) E(t) & 0 \\
      \mu_\pi (\Op{R}) E(t) & \Op{T} + V_{A^1\Sigma_u^+}(\Op{R}) -\hbar\omega_L &
      \sqrt{2} W_\mathrm{SO} \\
      0 & \sqrt{2}W_\mathrm{SO} &
      \Op{T} + V_{b^3\Pi_u}(\Op{R}) - W_\mathrm{SO} -\hbar\omega_L
    \end{pmatrix} \, ,      
  \end{equation}
\end{widetext}
where $\Op{T}$ denotes the kinetic energy operator and $V_i(\Op{R})$ the respective
potential energy curves. 
The scalar product between the transition dipole moment
and the polarization vector of the field is denoted as 
$\mu_{\sigma(\pi)}(\Op{R})$ for $\sigma$ ($\pi$) polarization. The laser field
is characterized by its temporal shape and central frequency,
$E(t)\exp(i\omega_L t)$.
The $\Op{R}$-dependence of the spin-orbit coupling $W_\mathrm{SO}$
is not known. It is therefore approximated
by its asymptotic value, $W_\mathrm{SO}=\mathrm{const}=1/3 \Delta E_\mathrm{FS}$,
which is given in terms of
the fine structure splitting, $\Delta E_\mathrm{FS}=237.5984$~cm$^{-1}$ for rubidium.
In Section~\ref{sec:SO}, two different model curves for $W_\mathrm{SO}(\Op{R})$
will be introduced, and the dependence of the results on
the specific description of the spin-orbit coupling 
will be discussed. 

For the $1_g$ and $0_g^-$ states, the spin-orbit coupling has not got resonant
character. In order to treat them on the same level of rigor as $0_u^+$, 
the spin-orbit interaction is nonetheless included explictly as in Eq.~(\ref{eq:Ham1}).
The Hamiltonians $\Op{H}_2$ and $\Op{H}_3$ then read
\begin{widetext}
  \begin{equation}
    \label{eq:Ham2}
    \Op{H}_2 =
    \begin{pmatrix}
      \Op{T} + V_{a^3\Sigma_u^+}(\Op{R}) & \mu_\sigma(\Op{R}) E(t) & 0 & \mu_\pi(\Op{R}) E(t) \\
      \mu_\sigma (\Op{R}) E(t) & \Op{T} + V_{^3\Pi_g}(\Op{R}) -\hbar\omega_L &
       W_\mathrm{SO} &W_\mathrm{SO} \\
      0 & W_\mathrm{SO} &
      \Op{T} + V_{^1\Pi_g}(\Op{R}) -\hbar\omega_L & -W_\mathrm{SO}\\
      \mu_\pi(\Op{R}) E(t) & W_\mathrm{SO} & -W_\mathrm{SO}&
      \Op{T} + V_{^3\Sigma^+_g}(\Op{R}) -\hbar\omega_L 
    \end{pmatrix} \, ,      
  \end{equation}
  and
%\end{widetext}
%\begin{widetext}
  \begin{equation}
    \label{eq:Ham3}
    \Op{H}_3 =
    \begin{pmatrix}
      \Op{T} + V_{a^3\Sigma_u^+}(\Op{R}) & \mu_\pi(\Op{R}) E(t) & \mu_\sigma(\Op{R}) E(t) \\
      \mu_\pi (\Op{R}) E(t) & \Op{T} + V_{^3\Sigma^+_g}(\Op{R}) -\hbar\omega_L &
      \sqrt{2} W_\mathrm{SO} \\
      \mu_\sigma(\Op{R}) E(t) & \sqrt{2} W_\mathrm{SO}&
      \Op{T} + V_{^3\Pi_g}(\Op{R}) -W_\mathrm{SO}  -\hbar\omega_L
    \end{pmatrix} \, .     
  \end{equation}
\end{widetext}
Both $1_g$ and $0_g^-$ states are coupled to the lowest triplet state
by $\sigma$- as well as $\pi$-polarization.
For simplicity the nuclear axis is chosen to be \textit{either} parallel or perpendicular
to the polarization axis of the laser field. For an ensemble of atoms in a
magneto-optical trap (MOT) with no preferred axis, an
average over all angles between the nuclear and polarization axes 
needs to be performed. This is beyond the scope of the current study.
Furthermore, rotational excitation is presently neglected and will be addressed
in future work.

\begin{figure}[tb]
  \centering
  \includegraphics[width=0.95\linewidth]{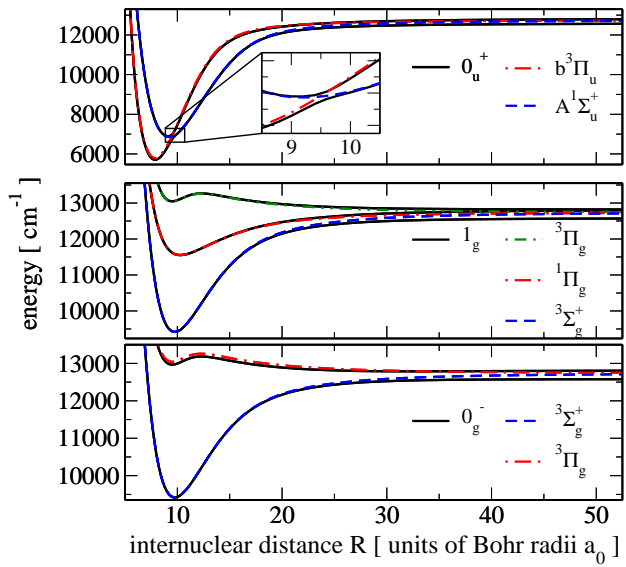}
  \caption{(Color online)
    The three potentials correlated to the $D_1$ line into which dipole allowed
    transitions may occur: $0_u^+$, $1_g$, and $0_g^-$ (black solid lines).
    Also shown are their counterparts which are correlated to the $D_2$ line and
    the corresponding Hund's case (a) potentials (red, blue and green
    dashed and dash-dotted lines).
  }
  \label{fig:pots}
\end{figure}

The potentials $V_i(\Op{R})$ have been obtained
by matching the results of \textit{ab initio} calculations~\cite{Olivier}
to the long-range dispersion potentials  %$V_\mathrm{asy}(\Op{R})=
$( C_3/\Op{R}^3 +) C_6/\Op{R}^6+C_8/\Op{R}^8$
(see Ref.~\cite{AymarJCP05} for details of the \textit{ab initio} calculations).
The coefficients for the $5S+5S$ asymptote are found in Ref.~\cite{MartePRL02}, 
while the coefficients for the $5S+5P$ asymptote are 
taken from  Ref.~\cite{GuterresPRA02}. The repulsive barrier of the 
($5S+5S$) potentials has been adjusted to give a triplet (singlet) 
scattering length of 99~a$_0$ (90~a$_0$). 
The excited state potentials in both Hund's case (a) and (c) representation
are shown in Fig.~\ref{fig:pots}.
Note that at long range the $0_g^-(P_{1/2})$ potential 
goes as $1/R^6$ since the $1/R^3$-terms cancel each other.
For the $1_g$ and  $0_g^-$ states, the spin-orbit coupling
influences mostly the long range part of the potentials while at short range
the singlet (triplet) character is retained. For the $0_u^+$ potentials, however,
resonant coupling between the $A^1\Sigma_u^+$ and $b^3\Pi_u$ states is observed which 
is due to the (avoided) crossing at $R\approx 9.5$~a$_0$
(see inset in Fig.~\ref{fig:pots}). The $0_u^+$
vibrational eigenfunctions therefore have mixed singlet/triplet character over
a large range of binding energies.
The $\Op{R}$-dependent transition dipole moments are also taken from the
\textit{ab initio} calculations~\cite{Olivier}.

\begin{figure}[tb]
  \centering
  \includegraphics[width=0.95\linewidth]{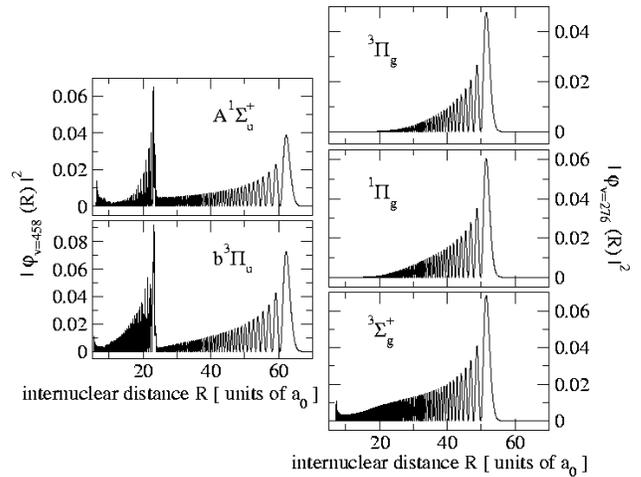}
  \caption{Vibrational wavefunctions of the $0_u^+$ (left) and $1_g$ (right) potentials
    with binding energies of 10.3~cm$^{-1}$ and 10.2~cm$^{-1}$ w.r.t. the $D_1$ line. }
  \label{fig:vibfcts}
\end{figure}

The Hamiltonians, Eqs.~(\ref{eq:Ham1}-\ref{eq:Ham3}),
are represented on a grid employing
a mapped grid method~\cite{SlavaJCP99,WillnerJCP04}. 
This allows to employ a fairly extended grid ($R_\mathrm{max}\approx 18500$~a$_0$)
with a comparatively small number of grid points ($N_\mathrm{grid}=1027$).
Such a large grid is needed to faithfully represent the scattering continuum
above the singlet ground and lowest triplet state potentials by box
states~\cite{ElianePRA04,ElianeEPJD04,My05}.
Diagonalization of the Hamiltonians, Eqs.~(\ref{eq:Ham1}-\ref{eq:Ham3}),
with $E(t)$ set to zero gives the vibrational energy levels and wavefunctions.
Fig.~\ref{fig:vibfcts} shows an example of these eigenfunctions of the
$0_u^+$ and $1_g$ states.
Some of the $0_u^+$ eigenfunctions are strongly perturbed by the resonant
coupling (shown on the left-hand side of Fig.~\ref{fig:vibfcts},
see also Ref.~\cite{SlavaJCP99})
as compared to the regular vibrational wavefunctions shown for the $1_g$ state.
Note that the peak at $R=25$~a.u. in the wavefunctions on the left-hand side of
Fig.~\ref{fig:vibfcts} corresponds to the outer turning point of the
$0_u^+(P_{3/2})$ potential.

The time-dependent Schr\"odinger equation,
\begin{equation}
  \label{eq:TDSE}
  i\hbar \frac{\partial}{\partial t}|\Psi(t)\rangle = \Op{H}(t) |\Psi(t)\rangle \,,
\end{equation}
is solved with a Chebychev propagator. 
Coherent effects resulting from laser pulses which overlap in time are not
of interest in the current context. 
Therefore the excitation and deexcitation steps are treated separately. 
In the first step,
an initial scattering state is excited by a PA (pump) pulse.
After a certain time delay a second (dump) pulse, suitably frequency 
shifted w.r.t. to the first one,
transfers the excited state population back to the electronic ground
state. Both laser pulses are assumed to have a Gaussian envelope and
possibly a frequency-chirp,
\begin{equation}
  \label{eq:pulse}
  E_j(t) = \frac{E_{0,j}}{\sqrt{f_j}} \Fkt{e}^{\frac{(t-t_j)^2}{2\sigma_j^2}}
  \Fkt{e}^{i\frac{\chi_j}{2}(t-t_j)^2}\; (j=P,D) \,.
\end{equation}
The Gaussian standard deviation $\sigma_{P(D)}$ is related to
the full width at half maximum (FWHM) of the intensity profile
$\tau_{P(D)}$ of the pump (dump) pulse
by $\tau_{P(D)}=2\sqrt{\ln 2}\sigma_{P(D)}$.
$t_\mathrm{P(D)}$ denotes the time at which the field amplitude is maximum,
$\chi_\mathrm{P(D)}$ is the time chirp of  the pump (dump) pulse. 
The stretch factor $f_\mathrm{P(D)}$ gives the ratio between
the pulse duration of the chirp to that of the corresponding transform-limited pulse,
$f=\tau^\mathrm{chirp}/\tau^\mathrm{tl}$ ($f_j=1$ for transform-limited pulses).
Some of the parameters characterizing a chirped pulse are related.
While chirping stretches the pulse in time, it leaves
its spectral bandwidth (FWHM) $\Delta\omega$ invariant,
$\Delta\omega = 2\sqrt{\ln 2}\Gamma$ with $\Gamma$ the Gaussian standard deviation
of the Fourier transform of $E_j(t)\Fkt{e}^{-i\omega_j t}$,
\begin{equation}
  \label{eq:spectrum}
  \tilde{E}_j(\omega) = E(\omega_j)
   \Fkt{e}^{\frac{(\omega-\omega_j)^2}{2\Gamma_j^2}}
   \Fkt{e}^{i\Phi_j\frac{(\omega-\omega_j)^2}{2}} 
\end{equation}
$ (j=P,D)$
and $\Gamma=2/\sigma^\mathrm{tl}$. Time and frequency domain
chirps are related to the spectral bandwidth via
\[
  \label{eq:chirps}
  \chi = \frac{\sqrt{f^2-1}}{f^2} \frac{\Gamma^2}{4} \,, \quad
  \Phi =   \sqrt{f^2-1} \frac{4}{\Gamma^2} \,.
\]
In the following, a pulse will be characterized  by its central frequency
$\omega_j$ (or, respectively, detuning from the $D_1$ line, $\Delta_j$),
FWHM $\tau_j$ of the transform-limited pulse (implying its spectral bandwidth
$\Delta\omega_j$), 
stretch factor $f_j$ which gives the strength of the chirp, the sign of the chirp and the
pulse energy. The latter is related to the maximum pulse amplitude $E_j$ and to the
duration $\sigma_j$ by
\[
\mathcal{E}_j = \eps_0 c \sqrt{\pi} \pi r^2 E_j^2 \sigma_j 
\]
with $\eps_0$ the electric constant, $c$ the speed of light and $r$ the
radius of the laser beam ($r=300\mu$m is assumed throughout this work).
These parameters are chosen to correspond to pulses which can be generated
from a femtosecond oscillator without amplification. This implies
in particular pulses with FWHM of up to 10~ps and
pulse energies of a few nano-Joule corresponding to peak intensities on the
order of 100 kW/cm$^2$.

\section{Excitation to $5S+5P_{1/2}$ states}
\label{sec:excitation}

The photoassociation of two free atoms to molecular levels of 
the  $0_u^+$, $1_g$, and $0_g^-$
states  is first studied. 
The continuum of scattering states
is represented by a finite number of box states~\cite{ElianePRA04,ElianeEPJD04,My05}.
These box states are normalized to one. To obtain the 
energy normalization which is usually employed for continuum states,
they have to be weighted by the density of states of
the box (see Ref.~\cite{ElianeEPJD04} for details).
The initial state of the calculations is chosen to be the 
s-wave scattering state
with collision energy $E_\mathsf{coll}$ equivalent to a temperature 
$T=E_\mathsf{coll}/k_B=105$~$\mu$K (with $k_B$ the Boltzmann constant).
Such a state
is typical for the conditions in a MOT. At these low temperatures
the short-range part of the scattering states is independent of temperature. 
This means that the position of the nodes of the scattering wavefunctions
approximately match the nodes of the last bound level.
If the Franck-Condon radius  corresponding to the central frequency of the pulse,
$R_C(\omega_L)$,
is located within this region determined by the extension of the last bound
level, the excitation probability is roughly temperature-independent.
In the following, $\pi$-polarization of the laser will be assumed, and
the rotational angular momentum of the molecules is taken to be $J=0$. 

\subsection{Excitation efficiency to the $0_u^+$, $1_g$, and $0_g^-$ states}
\label{subsec:efficiency}

\begin{figure}[tb]
  \centering
  \includegraphics[width=0.95\linewidth]{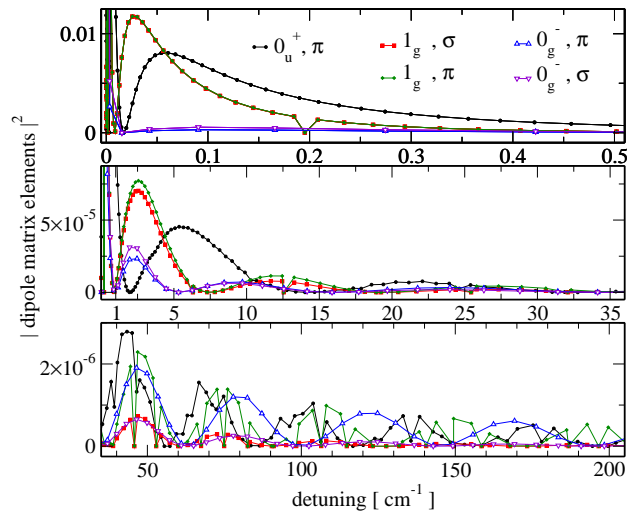}
  \caption{(Color online)
    Dipole matrix elements (absolute values squared)
    between an initial scattering state of
    two atoms with collision energy corresponding
    to T=105~$\mu$K and excited state vibrational levels
    vs. binding energy equal to the photoassociation laser
    detuning.
    The scattering state is normalized to one (see text for further explanations).
    Different electronic states are compared (the required  polarization 
    is denoted in the legend).
    \textit{The dpms are similar
      for transitions to $0_u^+$ and $1_g$ for all
      detunings. For $0_g^-$, they
      become comparable to those of the $0_u^+$ and $1_g$ states
      for detunings $\Delta_P > 1$~cm$^{-1}$.      
    }
  }
  \label{fig:FC}
\end{figure}

For weak fields, the excitation efficiency is completely determined by the
dipole matrix elements and by the detuning and spectral bandwidth of the laser.
Fig.~\ref{fig:FC} shows the absolute value squared of the
dipole matrix elements 
$|\langle \varphi^{exc}_{v'} |\mu_\pi(\Op{R})| \varphi^g_{T=105\mu\mathrm{K}} \rangle|^2$
(dpms) vs. binding energy $E^{exc}_{v'}$  or, respectively,
the laser detuning, for three different ranges of detunings ($exc=0_u^+, 1_g,0_g^-$). 
Very close to the atomic resonance ($\Delta_P<1$cm$^{-1}$, upper panel), 
the dpms for transitions into $0_g^-$ are about two orders of magnitude smaller
than those for transitions into $0_u^+$ and $1_g$. This is due to the $1/R^6$-nature
of $0_g^-$ at long range as compared to  $1/R^3$ for $0_u^+$ and $1_g$.
It is noteworthy that this region of small detunings where PA is most
efficient is \textit{un}likely to be accessible in an experiment with short pulses due
to the bandwidth of the pulse. Such pulses contain spectral components
corresponding to the atomic resonance, which can destroy the MOT ~\cite{Salzmann,Brown}. 
The pulse therefore has to be either far-detuned from the
atomic line or the resonant spectral components have to be filtered out. This
filtering can only be done with a finite spectral resolution which is of the
order of 1~cm$^{-1}$.
For larger detunings ($\Delta_P>1$cm$^{-1}$, Fig.~\ref{fig:FC}, middle and lower panel)
the dpms for all three potentials are of the same order of magnitude. The difference between
$1/R^3$- and  $1/R^6$-potentials lies only in the different density of molecular
levels, with that of a $1/R^6$-potential being significantly smaller. Since for
PA with \textit{pulses}, more than one level is resonant within
the spectral bandwidth, this density of states becomes important for the 
excitation efficiency. The overall excitation efficiency for
PA with short pulses will be smaller than one might
initially expect due to the inaccessibility of the range of very small detunings.
It has furthermore to be concluded from Fig.~\ref{fig:FC} that excitation by
pulses is unlikely to be selective: Only very few detunings exist where the dpms
are large for one potential and negligible for the other ones
(cf. $\Delta_P\sim 7.5$cm$^{-1}$, Fig.~\ref{fig:FC}, middle panel). In most cases,
the bandwidth of the pulse will comprise resonances with significant dpms for more than
one potential.

Table~\ref{tab:pexc} presents PA probabilities obtained from solving
the time-dependent Schr\"odinger equation, Eq.~(\ref{eq:TDSE}),
for all three Hamiltonians, Eqs.~(\ref{eq:Ham1}-\ref{eq:Ham3}), and
for transform-limited pulses.
\begin{table}[tb]
  \begin{ruledtabular}
    \begin{tabular}{cccc}
      State & $P_\mathrm{exc}$ &
      dpms ($E_{v'} = \hbar\Delta_P$)\footnote{$\Delta_P=4.1$~cm$^{-1}$, $\tau_P=10$~ps} &
      $P_\mathrm{low, last}$ \\ \hline
      $0_u^+$ & $2.9\times 10^{-5}$ & $3.5\times 10^{-5}$ & $5.3\times 10^{-5}$ \\
      $1_g$ & $3.8\times 10^{-5}$ & $4.7\times 10^{-5}$  & $1.0\times 10^{-4}$ \\ \hline\hline
      State & $P_\mathrm{exc}$ &
      dpms ($E_{v'} = \hbar\Delta_P$)\footnote{$\Delta_P=8.6$~cm$^{-1}$, $\tau_P=5$~ps} &
      $P_\mathrm{low, last}$ \\ \hline
      $0_u^+$ & $3.1\times 10^{-5}$ & $2.5\times 10^{-5}$ & $5.8\times 10^{-6}$ \\
      $0_g^-$ & $2.4\times 10^{-6}$ & $6.8\times 10^{-6}$ & $3.5\times 10^{-8}$ \footnote{%
        For $0_g^-$, the highest population in a bound level of the lowest triplet state
        is found for $v=$last-2, $P_\mathrm{g, last-2}=6.2\times 10^{-7}$ rather than
        $v=$last.} 
      \\ 
    \end{tabular}
  \caption{Excited state population after the pulse ($P_\mathrm{exc}$) compared to
    the absolute value squared of the dipole matrix element (dpms)
    between the initial state and one exemplary
    vibrational level (the one which is resonant with the
    central frequency of the pulse) for transitions from the singlet ground ($0_u^+$) and
    lowest triplet ($1_g$, $0_g^-$) state.
    The last column lists the population of the last bound \textit{ground}
    or lower triplet state level, respectively, 
    after the pulse. The pulse energy is 4.2 nJ in all cases,
    and $\pi$-polarization has been assumed.
  }
  \label{tab:pexc}
  \end{ruledtabular}
\end{table}
Due to the normalization of the box state, the PA probability is given
by the excited state population after the pulse, $P_\mathrm{exc}$. Each of the
detunings was chosen to be resonant with levels in both compared potentials. 
The corresponding dpms of the central level
is given in the third column of Table~\ref{tab:pexc}.
The PA probabilities for transitions to $0_u^+$ and $1_g$ at $\Delta_P=4.1$~cm$^{-1}$
are comparable and reflect the dpms of the central level. At $\Delta_P=8.6$~cm$^{-1}$,
the PA probability for  $0_u^+$ is an order of magnitude larger than for
$0_g^-$, while the corresponding dpms differ only by a factor of 3.7. This reflects
the different densities of states in a $1/R^3$ and a $1/R^6$ potential.
The last column of Table~\ref{tab:pexc} lists
the probability of forming molecules 
in the last bound level of the singlet ground or lowest triplet state.
For small detuning ($\Delta_P=4.1$~cm$^{-1}$), this
probability is even higher than that of creating excited state molecules, a
phenomenon which has already been observed for cesium~\cite{ElianePRA04}.
For larger detuning ($\Delta_P=8.6$~cm$^{-1}$), this probability is decreased
as compared to $P_\mathrm{exc}$ but is still significant. The explanation of this
behavior will be given below.

\begin{figure}[tb]
  \centering
  \includegraphics[width=0.95\linewidth]{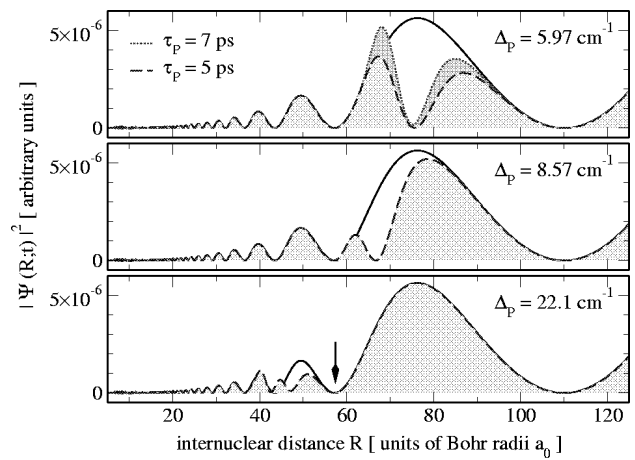}
  \caption{(Color online)
    The singlet ground state wavefunction (absolute value squared)
    before (black solid line)
    and after the pulse
    (colored dashed and dotted lines).
    The arrow indicates the position of the last node of the last bound ground state level.
    %The pulse energy is 4.2 nJ in all cases.
    Different detunings and pulse
    bandwidths are compared:
    \textit{The spectral width of the pulse determines the size of the
      photoassociation window
      while the detuning fixes its position.}
  }
  \label{fig:phig_tfinal}
\end{figure}

\begin{figure}[tb]
  \centering
  \includegraphics[width=0.95\linewidth]{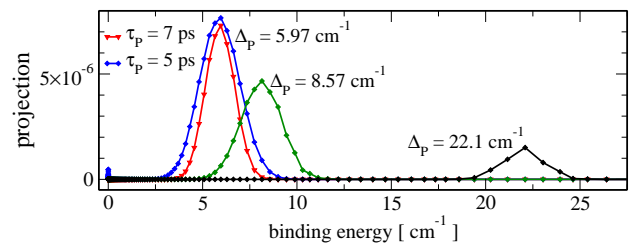}
  \caption{(Color online)
    The projection of the excited state wavefunction after the pulse
    onto the $0_u^+$ excited state
    vibrational eigenfunctions (absolute values squared)
    vs. the binding energy of the vibrational levels:
    \textit{The excited state
      vibrational distributions reflect the detuning from the atomic line
      (peak position) and the spectral width (peak width) of the pulse}.
%    (same parameters as in Fig.~\ref{fig:phig_tfinal}).
  }
  \label{fig:proje_tfinal}
\end{figure}

The PA pulse creates a wavepacket composed of a superposition of several vibrational
levels with time-dependent coefficients, on the excited electronic state. 
In addition the PA leaves a corresponding  'hole' in the ground
state wavefunction. This phenomena is illustrated in Figs.~\ref{fig:phig_tfinal} and
\ref{fig:proje_tfinal} for excitation from the singlet ground state to
$0_u^+$.
The ground state components before and after the pulse ($t=t_P\pm 4\sigma_P$)
are compared in Fig.~\ref{fig:phig_tfinal} for 
different detunings and pulse durations corresponding to different spectral
bandwidths. For $\Delta_P=5.97$~cm$^{-1}$,
the Franck-Condon point of the central frequency correponds to a 
maximum in amplitude of the ground state scattering
wavefunction. It also coincides
with the outermost maximum of the last bound level. The pulse cuts
a hole around the position of this maximum:  Population transfer occurs
within the \textit{photoassociation window}, defined by the
range of distances given by the Franck-Condon points $R_C(\omega)$
for all frequencies within the
bandwidth of the pulse~\cite{ElianePRA04}. The depth of the
hole is determined by the intensity or pulse energy. The width of the PA window
is given by the spectral bandwidth of the pulse, cf. the blue dashed and red dotted lines
in the upper panel of Fig.~\ref{fig:phig_tfinal} ($\Delta\omega_P=2.94$~cm$^{-1}$
for $\tau_P=5$~ps, $\Delta\omega_P=2.10$~cm$^{-1}$ for $\tau_P=7$~ps).
Farther detuning of the central frequency from the atomic line shifts the PA window
to shorter internuclear distances  (Fig.~\ref{fig:phig_tfinal},
middle and lower panels). At these shorter distances, the probability density of the
ground state wavefunction is significantly smaller, and hence less population
can be excited.

Fig.~\ref{fig:phig_tfinal} explains the relation between the PA detuning 
and the population of the last bound ground state level, $P_\mathrm{g,last}$,
for $0_u^+$. As reported in Table~\ref{tab:pexc}, $P_\mathrm{g,last}$ 
is larger than $P_\mathrm{exc}$ for $\Delta_P=4.1$~cm$^{-1}$,
but smaller  than $P_\mathrm{exc}$ for $\Delta_P=8.6$~cm$^{-1}$.
This is because at 4.1~cm$^{-1}$ the center of the PA window is close
to the position of the maximum of the ground state scattering
wavefunction, as seen similarly in the upper panel. Since  this position
is almost identical with the position of the outermost maximum of the last bound
level, and since this outermost maximum contains about 85\% of the probability
density of this level, population transfer into this level
via Rabi cycling is extremely efficient.
At 8.6~cm$^{-1}$, the PA window is moved toward the position of the last node
of the last bound level (cf. Fig~\ref{fig:phig_tfinal}, middle panel), where less
population is available to be transferred. In conclusion,
the PA pulse is most efficient with respect to both $P_\mathrm{exc}$ and
$P_\mathrm{g,last}$
if the Franck-Condon point of its central frequency,
$R_C(\omega_L)$, corresponds to the position
of the last maximum of the last bound level, and its spectral bandwidth is large
enough such that the PA window comprises this last peak.

Fig.~\ref{fig:proje_tfinal} displays the projections of the excited state wavepacket
generated by the pulse onto the vibrational levels of $0_u^+$. 
The Gaussian peaks reflect the Gaussian shape of the pulse envelope.
The peak position is determined by the detuning of the laser, while the
peak width mirrors the spectral bandwidth. The height of the peaks is entailed
by the dpms which decreases for larger detuning from the atomic
line (all pulses have the same pulse energy). Note the non-zero values of the
projection close to the atomic line for $\tau_P=5$~ps and $\Delta_P=5.97$~cm$^{-1}$.
Even though the spectral components close to $\Delta_P=0$ are extremely small,
the atomic resonance is excited due to its huge dpms. This
emphasizes the importance of filtering out spectral components around the atomic line.

\subsection{Positive frequency chirp: Enhanced excitation probability}
\label{subsec:chirp+}

The next step is to relax the condition of transform-limited Gaussian pulses
and to introduce a frequency chirp where the
central frequency of the pulse changes linearly with time. Chirped pulses
have been introduced in the context of molecular $\pi$-pulses~\cite{CaoPRL00}.
The principle underlying this concept is that if the pulse is much shorter than
the vibrational period, the nuclear motion is not resolved by the pulse.
The molecule can be treated as an effective two-level system, and 
chirping the pulse enforces adiabatic following conditions.
PA with chirped pulses has been discussed within a Landau-Zener picture. 
It was found that a \textit{positive} chirp maximizes population transfer by
countering the slope of the excited state potential~\cite{JiriPRA00,WrightPRL05}.

\begin{figure}[tb]
  \includegraphics[width=0.95\linewidth]{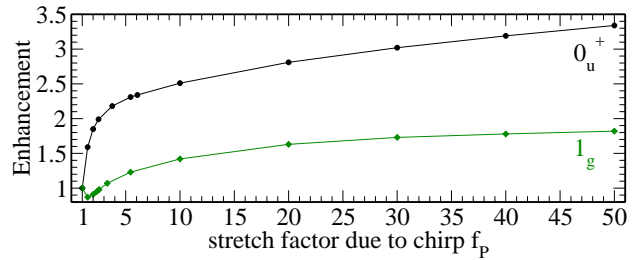}
  \caption{(Color online)
    Enhancement, i.e. ratio of excited state population after a positively chirped pulse
    to excited state population after the transform-limited pulse with the
    same spectral width $\Delta\omega_P$ vs. stretch factor $f_P$
    characterizing the strength of the chirp:
    \textit{A positive frequency chirp leads to more efficient population transfer for
    transitions to both $0_u^+$ and $1_g$ states.}
%    (The pulse energy is kept constant at 4.2 nJ and $\pi$-polarization has been assumed.)
    }
  \label{fig:chirp+}
\end{figure}

The calculations presented in 
Fig.~\ref{fig:chirp+} show the ratio of the excited
state population after a  positively chirped pulse to the 
excited state population after the corresponding transform-limited pulse
for transitions into $0_u^+$ and $1_g$. The pulse has a detuning of $\Delta_P=4.1$~cm$^{-1}$,
a pulse energy of 4.2~nJ, FWHM of 10~ps and a corresponding spectral bandwidth
of $\Delta\omega_P=1.47$~cm$^{-1}$.
The difference between these calculations and those of Refs.~\cite{JiriPRA00,WrightPRL05}
lies in the description of the initial state: For the Gaussian wavepacket of
Refs.~\cite{JiriPRA00,WrightPRL05} complete population inversion to
bound excited state levels can be obtained, while for a scattering state as in
Fig.~\ref{fig:phig_tfinal} population is transferred only from within the
photoassociation window (see Ref.~\cite{ElianeEPJD04} for a detailed discussion).
The larger enhancement for  $0_u^+$ than for $1_g$ is explained by the different topology
of the potentials leading to different values $R_c(\omega_L)$.
For $\Delta_P=4.1$~cm$^{-1}$, the center of the PA window is found
at $R_C=86.5$~a.u. for $0_u^+$, but at $R_C=69.9$~a.u. for $1_g$, i.e. for $0_u^+$
it is close to a maximum, while for $1_g$ it is close to
a node of the ground state wavefunction (cf. Fig.~\ref{fig:phig_tfinal}). 
Therefore within the PA window,
less population is available for transfer to $1_g$ than to $0_u^+$. A large amount of
this little population is already excited by a transform-limited
pulse of 4.2~nJ, thus chirping cannot increase it further.

\subsection{Negative frequency chirp: Shaping the excited state wavepacket}
\label{subsec:chirp-}

When the chirp is negative, large frequencies precede small ones during the pulse.
Consequently, the partial wavepackets at large distances are excited before the ones
at shorter distances. In an eigenstate-picture, levels with
large vibrational period are excited before those with smaller vibrational period.
The chirp can now be chosen to synchronize the excitation of partial wavepackets 
such that they will all arrive at exactly the same time at the inner classical
turning point~\cite{ElianePRA04,ElianeEPJD04}. The value of the chirp can be estimated
in terms of the vibrational period and the revival period of the level which is
resonant with the central frequency of the pulse, i.e. in terms of the
vibrational spectrum~\cite{ElianeEPJD04}.

\begin{figure}[tb]
  \includegraphics[width=0.95\linewidth]{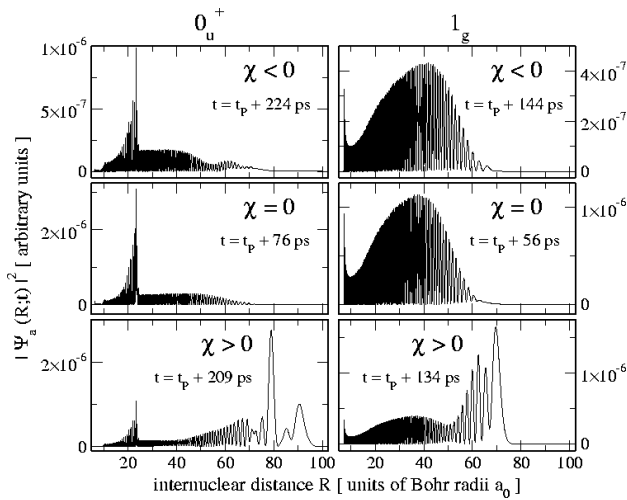}
  \caption{
    The best focussed wavefunctions for excitation to $0_u^+$ (left) and $1_g$ (right)
    for transform-limited 10~ps pulses (middle) and negatively (top) and
    positively (bottom)
    chirped pulses. The $A^1\Sigma_u^+$ (left) and $^3\Pi_g$ (right) components of the
    excited state wavefunctions are shown, i.e. the component of the state
    which is coupled
    to the singlet ground / lowest triplet state by the laser light:
    \textit{The resonant spin-orbit coupling in the $0_u^+$ states leads to a large
    population at short distances ($R\sim 20$~au).}
    $t_P$ denotes the maximum of the field amplitude,
    the stretch factors are $f_P =6.1$ (left) and $f_P=5.5$ (right),
    $\Delta_P=4.1$~cm$^{-1}$, $\mathcal{E}_P=4.2$~nJ.
  }
  \label{fig:psichirp-}
\end{figure}

Fig.~\ref{fig:psichirp-} shows how a negatively chirped pulse suppresses the
dispersion of the wavepacket while a positively chirped pulse enhances the dispersion.
A similar behavior has been reported for cesium in Refs.~\cite{ElianePRA04,ElianeEPJD04}.
The focussing effect for a negative chirp compared to a transform-limited pulse
is less pronounced than reported in
Refs.~\cite{ElianePRA04,ElianeEPJD04}. This is
due to the larger detuning: Close to the dissociation limit, the vibrational periods
increase exponentially and
more correcting for dispersion is needed. Therefore active shaping of the
excited state wavepacket by a negative chirp becomes less crucial with increasing
detuning.

A negative chirp compared to a
transform-limited pulse can suppress the  PA probability. In the range of
detunings of a few wavenumbers, negative chirp either left the PA
probability constant or reduced it by up to a factor of ten.
The reduction in PA probability was observed
for both $0_u^+$ and $1_g$ potentials. This is unlike 
the case of cesium where a negative chirp lead to
an increase of the PA probability by a factor of three~\cite{ElianePRA04}. 

The resonant spin-orbit coupling in the case of the $0_u^+$-potentials~%
\footnote{Note that the optimal chirp needs to be estimated from the Hund's case (a)
$A^1\Sigma^+_u$ potential without spin-orbit coupling.
The vibrational spectrum is so strongly perturbed
by the resonant coupling that the chirp as a function of binding energy exhibits
divergences.}
causes a piling-up of population around $R=25$~a.u. which corresponds to the outer
turning point of the $0_u^+(P_{3/2})$ potential. It is in line 
with the shape of the eigenfunctions (cf. the left-hand sides of Figs.~\ref{fig:psichirp-}
and \ref{fig:vibfcts}). In the case of the $1_g$-potentials, the maximum of population
is found for $R=40$~a.u.
In a pump-dump scheme, a second pulse  transfers these wavepackets back to the
singlet ground or lowest triplet state
with the goal of populating deeply bound levels. The positions
of the maxima, $R=25$~a.u. and $R=40$~a.u., therefore have to be compared to
the outer turning point of the bound ground state levels. Since levels with
binding energy larger than one wavenumber have their outer
turning point at distances shorter than $R=35$~a.u., 
efficient population transfer into these more deeply bound levels can only be
expected for $0_u^+$. Levels bound by less than 1~cm$^{-1}$ can be 
efficiently populated by  a single PA pulse and therefore do not require a
two-color scheme.

Finally, the population $P_\mathsf{g,last}$ of the last bound levels of the
singlet ground and lowest triplet states is 
roughly identical for  pulses with the same frequency content, i.e. it is independent
of a chirp.

\section{Deexcitation to $X^1\Sigma_g^+$ ground and
  $a^3\Sigma_u^+$ lower triplet state molecules}
\label{sec:deexcitation}

In order to obtain stable molecules in the singlet ground or lowest triplet states, 
a second or dump pulse may be applied to the excited state
wavepacket. Some vibrational levels in the  $X^1\Sigma_g^+$ ground and
  $a^3\Sigma_u^+$ lower triplet states are already populated by the first pulse.
  However, these levels are very loosely bound. A dump pulse can be optimized
  to populate more deeply bound levels. 
The concepts developed in Ref.~\cite{My05} will be
employed, in particular the time-dependence of the dpms
between the excited state wavepacket and the bound ground / lowest triplet state levels, 
and the idea of employing a narrow-bandwidth pulse,
suitably detuned w.r.t. to
the PA pulse, to achieve transfer into a single vibrational level.
As was shown in Ref.~\cite{My05}, for weak dump pulses
the population transfer to the lowest triplet state is completely determined
by the dpms and the central frequency and
spectral bandwidth of the pulse.
The difference between the current study and Ref.~\cite{My05} is due to
the stabilization mechanism allowing for efficient transfer to more deeply
bound levels. In the study on cesium, the stabilization mechanism was
most efficient for the wavepacket localized
at the soft repulsive wall of the outer well of the
$0_g^-(P_{3/2})$ state. Presently,
for rubidium below the $D_1$ line, stabilization is afforded 
by  vibrational levels of $0_u^+(P_{1/2})$ which are in resonance with 
a level of $0_u^+(P_{3/2})$.

\begin{figure}[tb]
  \includegraphics[width=0.95\linewidth]{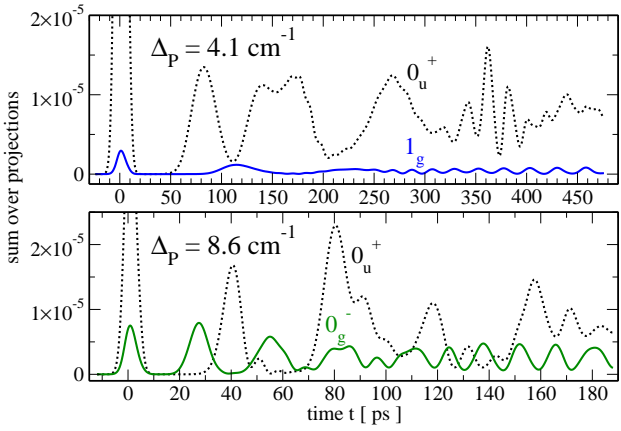}
  \caption{(Color online)
    The sum over all projections (absolute values squared)
    of the excited state wavepacket to bound
    singlet ground / lowest triplet state levels
    after excitation to $0_u^+$ (black dotted lines), 
    to $1_g$ (blue solid line, top), and to $0_g^-$ (green solid line, bottom):
    \textit{Deexcitation into bound levels is far more efficent from
    the $0_u^+$ state than from $1_g$ or $0_g^-$.}
    The transform-limited excitation pulses have
    $\tau_P=10$~ps (top),  $\tau_P=5$~ps (bottom) and pulse energy of 4.2~nJ.
    %$\pi$-polarization has been assumed.)
  }
  \label{fig:proje2gTL}
\end{figure}

Since coherent effects between the pump and the dump pulses are neglected,
the excited state population after the PA pulse can be normalized to one.
Populations correspond then directly to probabilities.
Fig.~\ref{fig:proje2gTL} shows the sum over the dpms
between the excited state wavepacket $\Psi_e(t)$
and all bound singlet ground / lowest triplet state levels,
$\sum_{v''} |\langle \varphi^g_{v''} |\mu_\pi(\Op{R})| \Psi_e(t) \rangle|^2$
for $0_u^+$, $1_g$, and $0_g^-$ ($e$ denotes the channel which is coupled
by the laser field to the singlet ground or lowest triplet state
in Eqs.~(\ref{eq:Ham1}-\ref{eq:Ham3}),
i.e. $A^1\Sigma_u^+$ and $^3\Sigma^+_g$). 
The sum over projections is compared for $0_u^+$ and $1_g$ at a PA pulse detuning
of $\Delta_P=4.1$~cm$^{-1}$, and for  $0_u^+$ and $0_g^-$ at  $\Delta_P=8.6$~cm$^{-1}$.
The origin of time is set equal to the time when the amplitude of the PA pulse is
maximum, $t_P=0$. The sum over projections
shows oscillations which reflect the excited state wavepacket dynamics.
At later times, the oscillations are washed out due to wavepacket dispersion
(the excitation pulses are transform-limited). At both detunings, the
projections are larger for $0_u^+$ than for $1_g$ and $0_g^-$, respectively, i.e.
stabilization is expected to be most efficient for $0_u^+$.

\begin{figure}[tb]
  \includegraphics[width=0.45\linewidth]{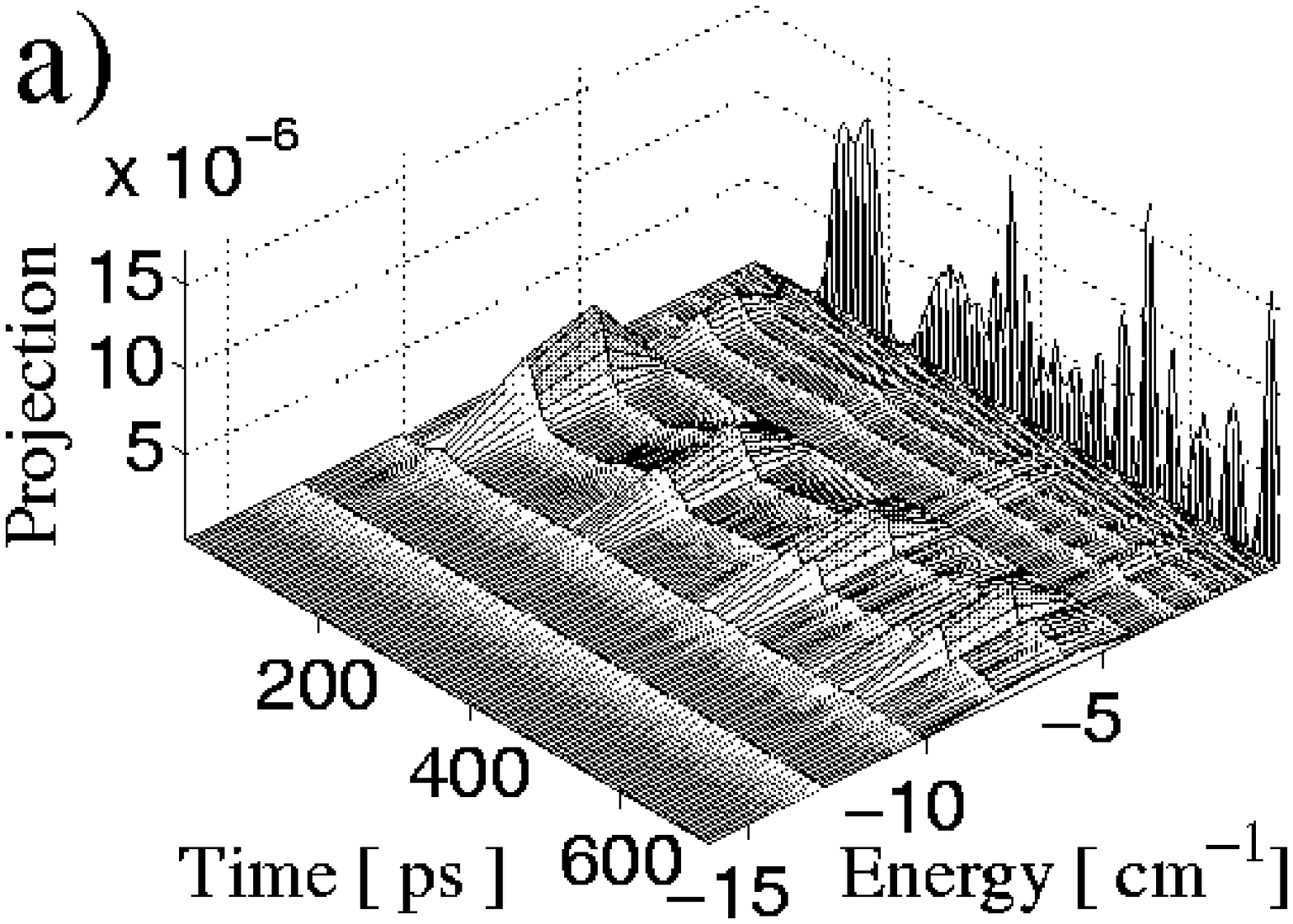}
  \includegraphics[width=0.45\linewidth]{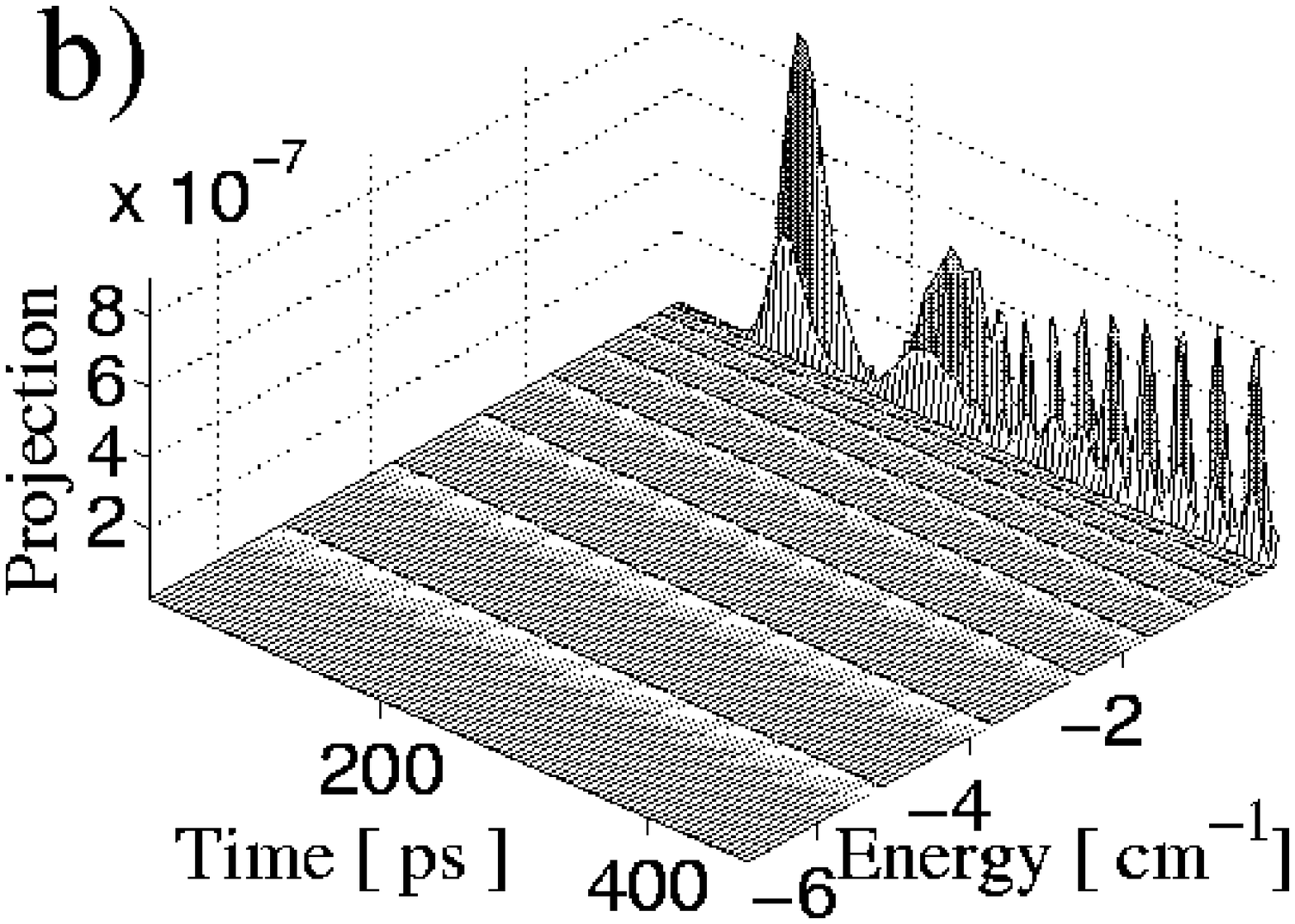}
  \includegraphics[width=0.45\linewidth]{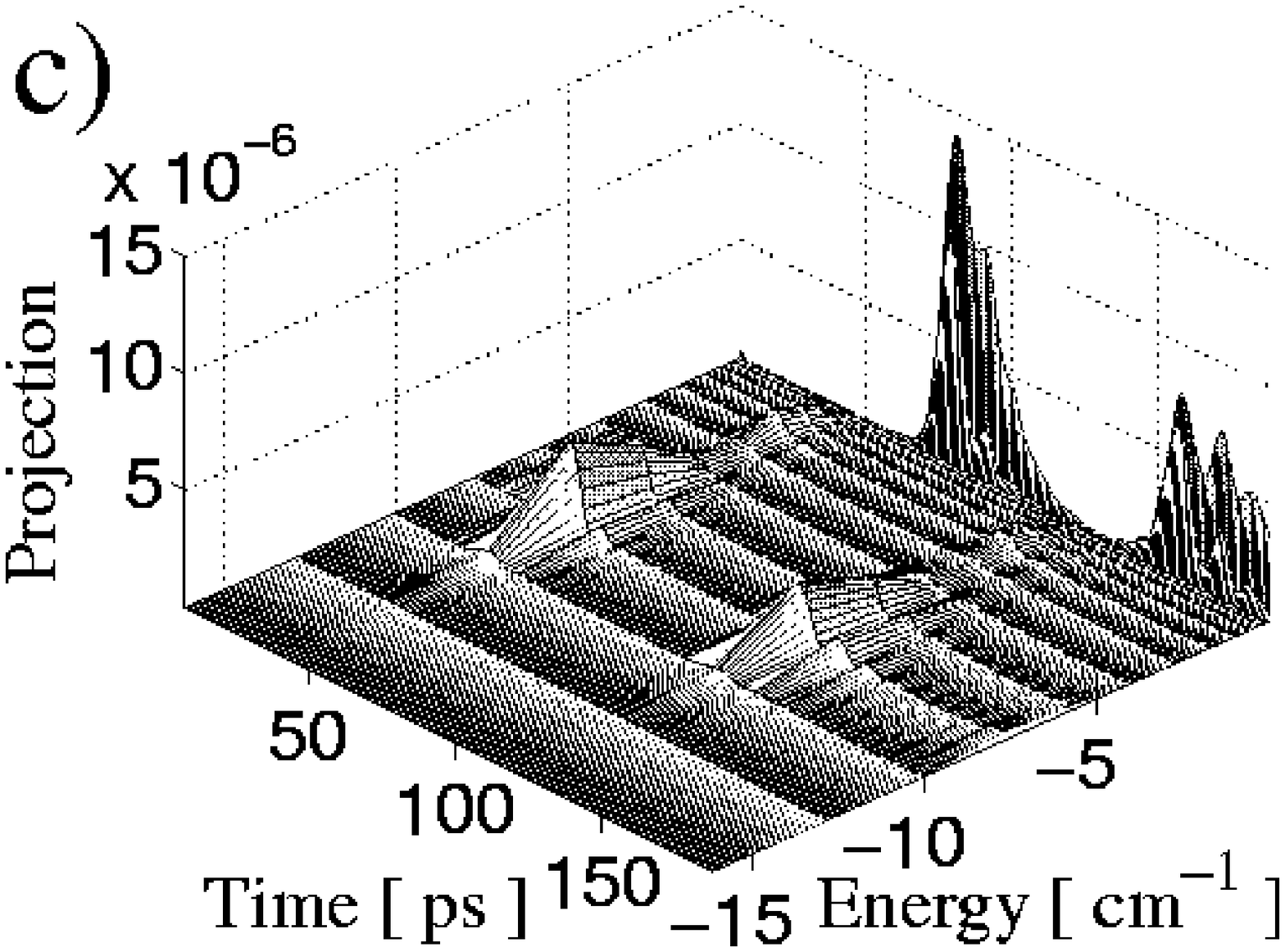}
  \includegraphics[width=0.45\linewidth]{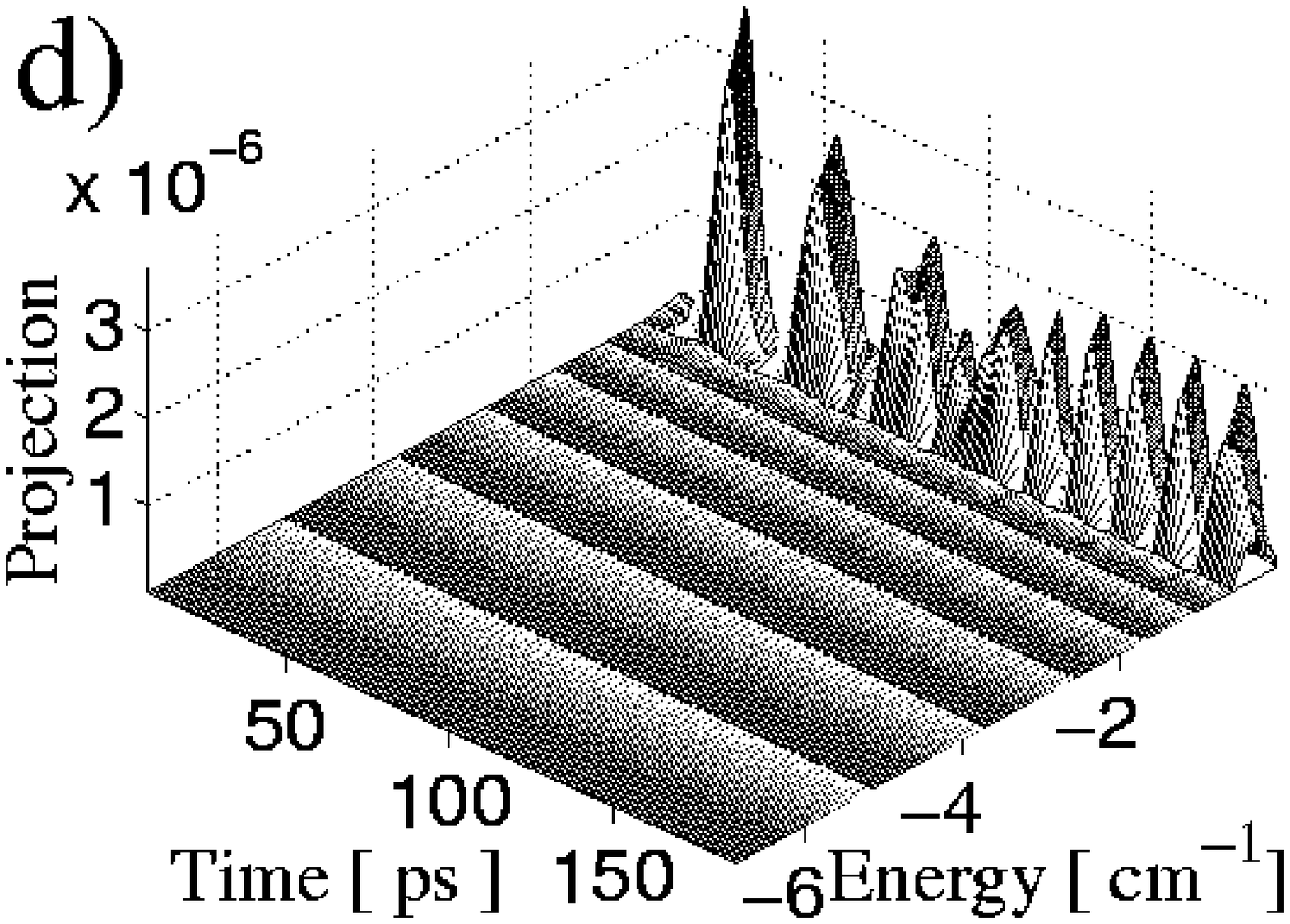}
  \caption{The projections (absolute values squared)
    of the excited state wavepacket to all bound
    singlet ground / lowest triplet state
    levels vs. time and binding energy after excitation to $0_u^+$ (a,c), $1_g$ (b),
    and $0_g^-$ (d) (a,b: $\Delta\omega_P=4.1$~cm$^{-1}$, c,d: $\Delta\omega_P=8.6$~cm$^{-1}$):
    \textit{Deexcitation from the $1_g$ and $0_g^-$ states can populate only 
    the last bound levels, while $0_u^+$ allows for transitions into more
    deeply bound ground state levels.}
  }
  \label{fig:ovlp}
\end{figure}

The sum over the dpms does not reveal which
levels in the singlet ground and lowest triplet states
are accessible by the dump pulse. 
Fig.~\ref{fig:ovlp} therefore shows the dpms,
$|\langle \varphi^g_{v''} |\mu_\pi(\Op{R})| \Psi_e(t) \rangle|^2$, as a function
of time and binding energy, $E_{v''}$, of the singlet ground / lowest
triplet state levels $v''$ (same 
parameters as in Fig.~\ref{fig:proje2gTL}).
For both $1_g$ and $0_g^-$ states, deexcitation can transfer population only into
the last two to three levels of the lowest triplet state (Fig.~\ref{fig:ovlp}~b, d).
These levels are bound by less than 0.1~cm$^{-1}$, and they are already
populated efficiently by the PA pulse (cf. Table~\ref{tab:pexc}).
In contrast, for $0_u^+$ dpms with both the last three and more deeply bound levels
are significant. The levels which are bound by 10~cm$^{-1}$ to 4~cm$^{-1}$
have vibrational index $v''=109$ to $v''=112$ (the last bound level has
$v''=120$). The outer classical turning points of the corresponding wavefunctions
are located between $R\sim24$~a.u.
and $R\sim28$~a.u. (as compared to  $R\sim170$~a.u. for the last bound level),
i.e. their outermost maximum is located at about the same position as the peak
in the excited state wavepackets in the left hand side of Fig.~\ref{fig:psichirp-}.
It can
therefore be concluded that the resonant spin-orbit coupling in the $0_u^+$
states provides the stabilization mechanism into
more deeply bound ground state levels.
It is obvious from 
Fig.~\ref{fig:ovlp}a and c that  the vibrational period
of the excited state wavepacket as well as its compactness is determined by the
detuning  as expected.

\begin{figure}[tb]
  \includegraphics[width=0.45\linewidth]{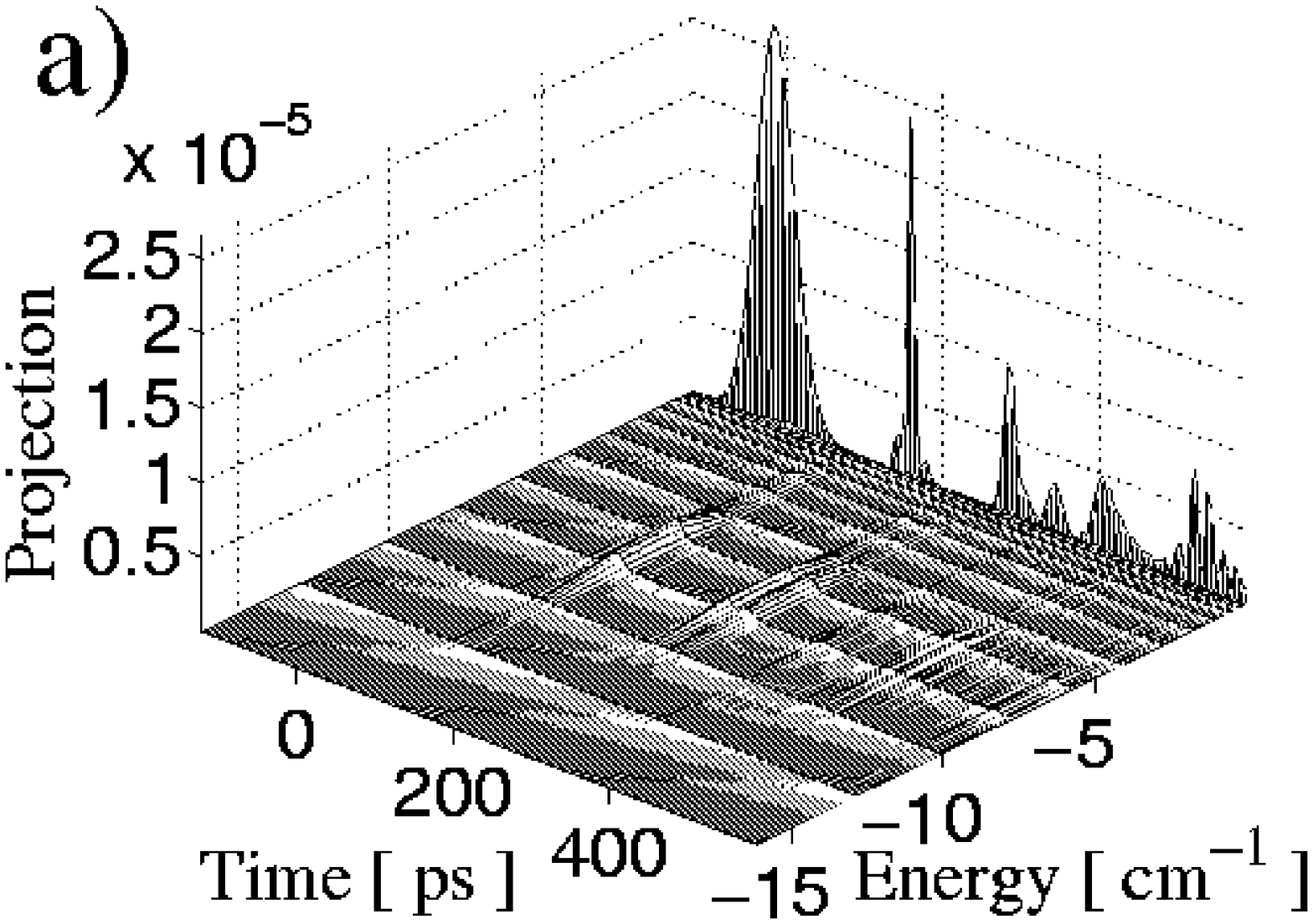}
  \includegraphics[width=0.45\linewidth]{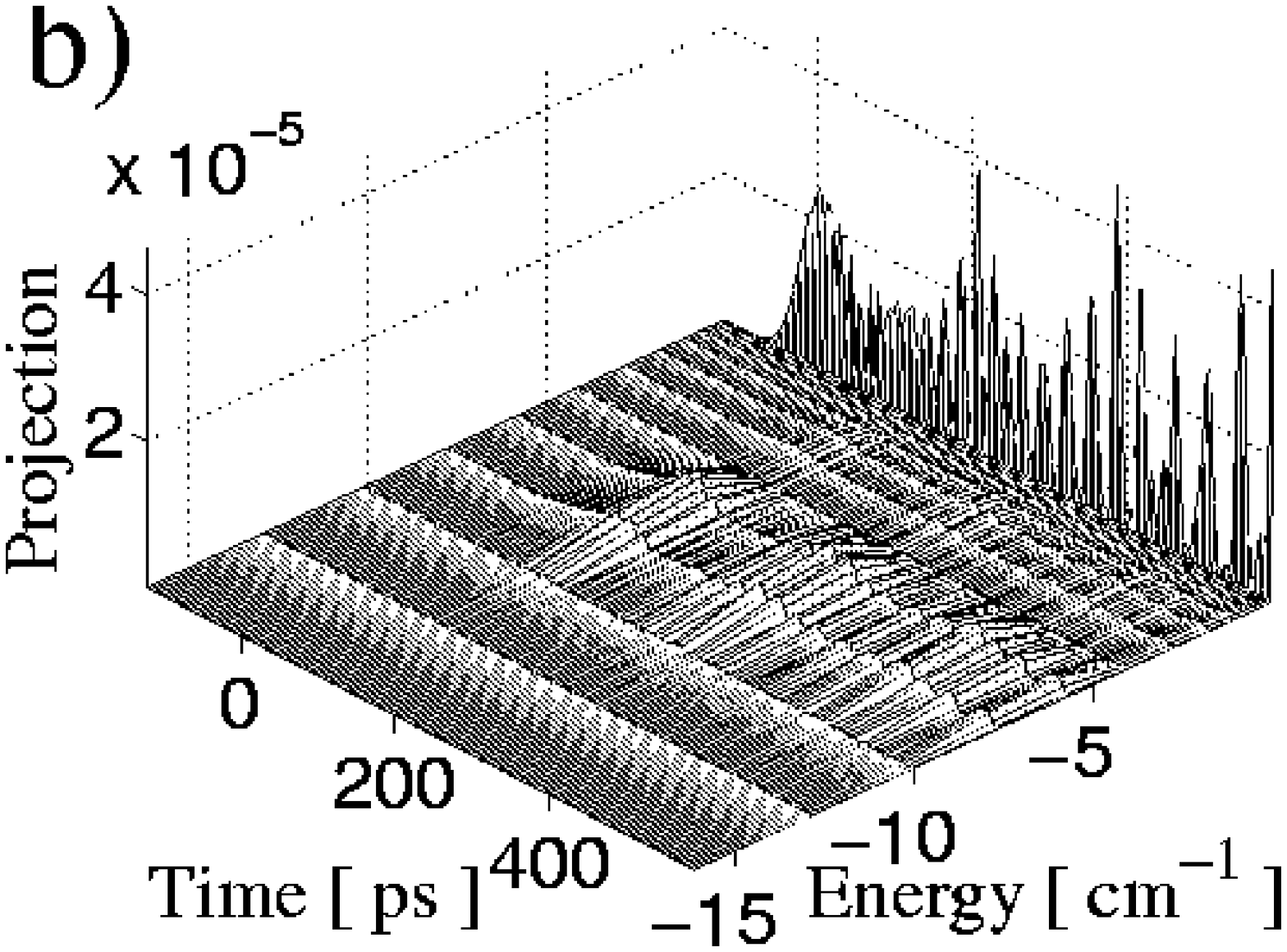}
  \caption{The projections (absolute values squared)
    of the excited state wavepacket to all bound ground state
    levels vs. time and binding energy after excitation by negatively (a) and positively
    (b) chirped pulses to $0_u^+$ ($f_P=6.1$, cf. the lhs of Fig.~\ref{fig:psichirp-}
    for examples of the excited state wavepackets and
    Fig.~\ref{fig:ovlp}a for the projections of the
    corresponding transform-limited case): \textit{
    Deexcitation into more
    deeply bound ground state levels is possible after transform-limited
    as well as after chirped pump pulses.}
     }
  \label{fig:ovlpchirp}
\end{figure}

The dpms of the $0_u^+$ excited state wavepacket after excitation with
chirped PA pulses is shown in Fig.~\ref{fig:ovlpchirp}.
A negative frequency chirp of the excitation pulse leads to a compact wavepacket
oscillating in the
excitated state potential. These oscillations are reflected in the clearly separated
peaks of the projections (Fig.~\ref{fig:ovlpchirp}a).
After a positively chirped excitation pulse the wave
packet is spread out in coordinate space and the peaks of the projections are smeared out
(Fig.~\ref{fig:ovlpchirp}b). Since chirping does not affect the spectral bandwidth
of the pulse, the excited state wavepackets after chirped and after transform-limited
pulses are composed of the same vibrational levels. Hence the ground state levels which
are accessible by a second pulse are identical
(cf. Fig.~\ref{fig:ovlpchirp} and Fig.~\ref{fig:ovlp}a).

\begin{figure}[tb]
  \includegraphics[width=0.9\linewidth]{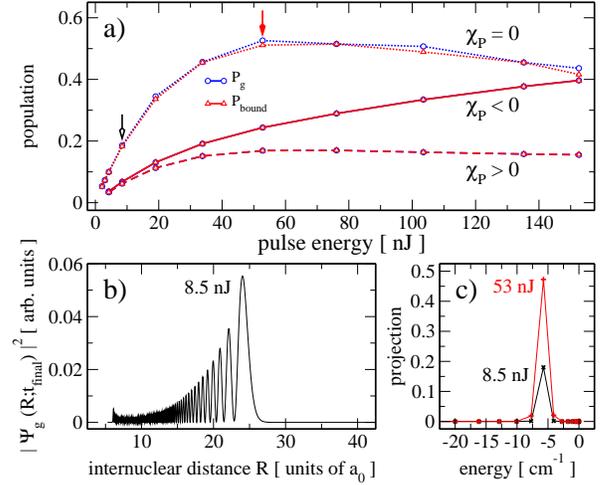}
  \caption{(Color online)
    a) Ground state population $P_g$ (blue circles) and population of bound ground state
    levels $P_\mathrm{bound}$ (red triangles)
    after a transform-limited dump pulse with FWHM of 10 ps (solid lines)
    following excitation 
    by a transform-limited (dotted lines), a negatively (solid lines) and a positively
    (dashed lines) chirped pump pulse.
    b) The final ground state wavepacket after a dump pulse with an energy of 8.5 nJ
    (corresponding to the black arrow in a).
    c) The projection of the final ground state wavepacket onto vibrational levels
    (absolute values squared), i.e. the
    vibrational distribution vs. binding energy after dump pulses with pulse
    energies of 8.5 nJ and 53 nJ (corresponding to the black and red arrows in a).
    \textit{Up to 50\% of the excited state wavepacket can be transferred to a single,
    more deeply bound ground state level.}
  }
  \label{fig:dump_TL}
\end{figure}

The outcome of the calculations with a second or dump pulse are now presented. 
The goal is to transfer as much of the excited state wavepacket  as possible  
into a single, more deeply bound ground state level. 
Calculations have therefore been done only for the $0_u^+$ states.
The initial state of the example presented in the following is
given by the excited state wavepacket after
excitation by a PA pulse with $\Delta_P=4.1$~cm$^{-1}$ and $\tau_P=10$~ps
(cf. left-hand side of Fig.~\ref{fig:psichirp-}).
The time-delay between PA and stabilization pulses which best
achieves the goal can be read off the time-dependent dpms (cf. Figs.~\ref{fig:ovlp} and
\ref{fig:ovlpchirp}, for example $t_D-t_P=81.5$~ps after the transform-limited PA pulse). 
The spectral bandwidth of the dump pulse is determined
by the requirement of populating a single vibrational level, chosen to be $v''=111$.
$\Delta\omega_D$ needs to be smaller than the
vibrational level spacing which is approximately 1.7~cm$^{-1}$
for this level. This corresponds to dump pulses with FWHM $\tau_D\ge 8$~ps.
The detuning is chosen such that the energies of the excited state wavepacket and the
target level are brought into resonance, $\Delta_D=-1.64$~cm$^{-1}$ for $v''=111$,
i.e. the dump pulse is blue-detuned w.r.t. the $D_1$ line.
The ground state population, $P_g=|\langle g|\Psi(t)\rangle|^2$,
and the population of bound ground state levels,
$P_\mathrm{bound}=\sum_{v''}|\langle \varphi^g_{v''}|\Psi(t)\rangle|^2$, 
after a dump pulse of $\tau_D=10$~ps is shown in
Fig.~\ref{fig:dump_TL}a as a function of pulse energy.
Since the narrow-bandwidth dump pulses are resonant only with bound ground state levels and
not with the ground state continuum, $P_g$ and $P_\mathrm{bound}$ are basically identical.
The calculations of Fig.~\ref{fig:dump_TL}a start from three different initial states,
the excited wavepacket after a 10~ps transform-limited PA pulse (dotted lines) as well as
after the corresponding positively and negatively chirped
PA pulses (dashed and solid lines, respectively,
cf. Figs.~\ref{fig:ovlp}a and \ref{fig:ovlpchirp}).
The highest ground state population is achieved after a transform-limited PA pulse.
The positively chirped PA pulse provides the least effective starting point for stabilization.
This is due to the large wavepacket dispersion. Generally, the achieved ground state
population is much higher than expected from the dpms: Up to 50\% of the excited state
wavepacket can be transferred to the ground state. 
The vibrational distribution of the
final ground state wavepacket, $|\langle \varphi^g_{v''}|\Psi_g(t)\rangle|^2$, shown in 
Fig.~\ref{fig:dump_TL}c demonstrates that as intended this population  ends up almost
exclusively in a single vibrational level, namely $v''=111$ bound by 5.74~cm$^{-1}$.
An example of a ground state wavepacket  after the dump pulse
is shown in Fig.~\ref{fig:dump_TL}b which confirms the pure eigenstate nature of
$\Psi_g(R;t)$. The enhanced population transfer by the stabilization pulse
is attributed to a dynamical effect due to spin-orbit coupling: 
The excited state wavepacket has components on both
$A^1\Sigma_u^+$ and $b^3\Pi_u$ states, but only the $A^1\Sigma_u^+$ component is coupled
to the ground state by the field. When the dump pulse acts, the $A^1\Sigma_u^+$ component
is depleted by the pulse, but 'refilled' by the resonant spin-orbit coupling.
Thus significantly more population is channeled to the ground state
than expected from the dpms which
only contain the $A^1\Sigma_u^+$ component. This effect
is most pronounced when the wavepacket is focussed around the spin-orbit peak (cf.
Fig.~\ref{fig:psichirp-}, lhs) and therefore works best after transform-limited or
negatively chirped PA pulses.

Transform-limited dump pulses with FWHM of 5 ps give similar results with slightly broader
vibrational distributions. Chirping the
dump pulses decreases the population transfer into more deeply
bound levels: The chirp stretches
the pulses such that the duration becomes much larger than the time in which the
wavepacket stays at short distances.

\section{Sensitivity on the description of the spin-orbit coupling}
\label{sec:SO}

The spin-orbit coupling having resonant character for
the $0_u^+$ states provides a mechanism for the formation of ground state
molecules bound by a few wavenumbers.
The extent to which the dump step depends on the specific model
of the spin-orbit coupling (SOC) is explored.
Two  model curves~\cite{SlavaPRA00a,SlavaPRA00b} to describe the $R$-dependence
of the SOC are employed. At this time,
no \textit{ab initio} or spectroscopic data allowing
for a more accurate description are available.
The first model curve, $W_1(R)$, has been obtained from \textit{ab initio} calculations
of the $R$-dependent SOC of Cs$_2$~\cite{Spiess} and scaling with
the ratio of the fine-structure splittings $\Delta E_\mathrm{FS}$ of rubidium and
cesium~\cite{SlavaPRA00b}. $W_1(R)$ displays a minimum at $R\approx 11$~a.u., and
at the crossing point of the potentials (at $R\approx 9.5$~a.u.), 
the SOC is reduced to about 65\% of its asymptotic value.
It was argued in Ref.~\cite{SlavaPRA00b} that the reduction of the coupling at
the crossing point is probably overestimated in $W_1(R)$. Therefore a second model curve,
$W_2(R)$, was introduced showing a similar dependence on $R$ as $W_1(R)$ but
with the coupling at the crossing point reduced to only 
80\% of the asymptotic value. The $R$-dependence of
both functions, $W_1(R)$ and $W_2(R)$, is displayed
in Fig.~2 of Ref.~\cite{SlavaPRA00b}. At long range, both curves become constant and
equal the atomic value.

\begin{figure}[tb]
  \includegraphics[width=0.9\linewidth]{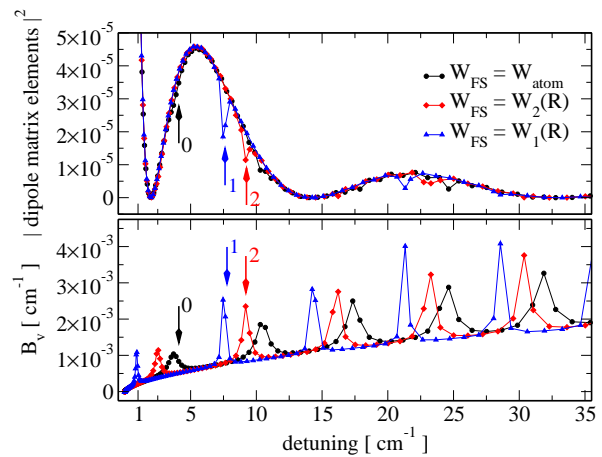}
  \caption{(Color online)
    The absolute values squared of the dipole matrix elements
    $|\langle \varphi^{exc}_{v'} |\mu_\pi(\Op{R})| \varphi^g_{T=105\mu\mathrm{K}} \rangle|^2$
    (top) and the rotational constants $B_{v'} = \langle 1/(2\mu \Op{R}^2)\rangle$
    of the $0_u^+$ excited states
    (bottom) for three different models of the spin-orbit coupling.
    The peaks of the rotational constants correspond to levels which are strongly
    perturbed by the resonant coupling.
    The specific model of the spin-orbit coupling has almost no influence on the
    dipole matrix elements. It does affect the position of the strongly perturbed levels
    which are crucial for deexcitation toward more deeply bound ground state molecules.
    \textit{Therefore the excitation probability is expected to be roughly independent
      of the description of the spin-orbit coupling. However, in order to assure
      an efficient dump step, the pump detuning
      needs to be adjusted such that resonant excited state levels are populated.
    }
  }
  \label{fig:Brot}
\end{figure}

In the following, the results
of Secs.~\ref{sec:excitation} and \ref{sec:deexcitation}
for $W_\mathrm{SO}= \mathrm{const}$ are compared to those obtained with
$W_\mathrm{SO}=W_1(R)$ and $W_\mathrm{SO}=W_2(R)$
for transitions via $0_u^+$.
The binding energies of the excited state levels are slightly shifted
for $W_1(R)$ and $W_2(R)$ as compared to constant coupling.
These shifts are negligible
with respect to the bandwidth of the pulse.
The excitation probability is determined by the dipole matrix
elements between the initial scattering state and bound vibrational
levels of the  $0_u^+$ excited states. The absolute values squared of
the dipole matrix elements (dpms) are shown in the upper panel
of Fig.~\ref{fig:Brot} for the three different SOC models.
Except for single levels where the wavefunction is strongly modified by
the coupling, the dpms are almost identical. This is to be
expected because the overlap between the scattering state and the excited
state levels is biggest at large distances $R$ where all three SOC
are constant.
For the levels where the dpms are visibly influenced, the value is reduced only by
a factor $\gtrsim 0.5$. The overall excitation probability should therefore
not significantly depend on the specific model of the SOC.
This is confirmed by inspection of
the excited state populations after the PA pulse, $P_\mathrm{exc}$, listed in
Table~\ref{tab:SO}.
\begin{table}[tb]
  \begin{ruledtabular}
    \begin{tabular}{cccccc}
      $\Delta_P$ & $W_\mathrm{SO}$ & $\tau_P$ & $P_\mathrm{exc}$ & dpms  &
      $P_\mathrm{g, last}$ \\
      $[\mathrm{cm}^{-1}]$  &                 & [ps]  & & ($E_{v'} = \hbar\Delta_P$) & \\
      \hline
      4.1 & const & 10    &$2.9\times 10^{-5}$ & $3.5\times 10^{-5}$ & $5.3\times 10^{-5}$ \\
      4.1 & $W_1(R)$ & 10 &$2.9\times 10^{-5}$ & $3.9\times 10^{-5}$  & $5.3\times 10^{-5}$ \\
      \hline
      7.6 & $W_1(R)$ & 10 &$1.6\times 10^{-5}$ & $2.2\times 10^{-5}$ & $9.2\times 10^{-6}$ \\
      7.6 & $W_1(R)$ & 5  &$4.6\times 10^{-5}$ & $2.2\times 10^{-5}$ & $9.2\times 10^{-6}$ \\
      \hline
      9.2 & $W_2(R)$ & 10 &$9.2\times 10^{-6}$ & $1.1\times 10^{-5}$ & $4.2\times 10^{-6}$ \\
      9.2 & $W_2(R)$ & 5  &$2.4\times 10^{-5}$ & $1.1\times 10^{-5}$ & $4.2\times 10^{-6}$ \\
    \end{tabular}
  \caption{Excited state population after the PA pulse ($P_\mathrm{exc}$)
    for transitions to $0_u^+$ and three different models of
    the spin-orbit coupling, $W_\mathrm{SO}$. Also listed are 
    the absolute values squared of the dipole matrix element (dpms)
    between the initial state and
    the vibrational level resonant with the
    central frequency, and
    the population of the last bound \textit{ground} state level
    after the pulse. The employed detunings are indicated by arrows in Fig.~\ref{fig:Brot}.
    The pulse energy is 4.2 nJ in all cases,
    and $\pi$-polarization has been assumed.
  }
  \label{tab:SO}
  \end{ruledtabular}
\end{table}

For deexcitation to more deeply bound molecular levels by the second pulse,
it is important for the PA pulse to
excite $0_u^+$ vibrational levels which are strongly perturbed.
The resonant or non-resonant character of the excited state wavefunctions
depends rather sensitively on the model of SOC.
Recalling Fig.~\ref{fig:vibfcts}, this is not surprising since the resonant
character is caused by the coupling at shorter distances (the outer turning point
of $0_u^+(P_{3/2})$ is $R\approx 20$ for binding
energies of a few cm$^{-1}$) where $W_1(R)$ and $W_2(R)$ differ from
$W_\mathrm{SO}=\mathrm{const}$.
Perturbed and regular vibrational levels can be differentiated by
their rotational constants $B_{v'}=<1/(2\mu \Op{R})>$ as discussed in detail in
Refs.~\cite{SlavaJCP99,SlavaPRA00a}. The rotational constants for the three different
SOC are therefore compared
in the lower panel of Fig.~\ref{fig:Brot}.
The peaks which are superimposed on the smooth dependence of the $B_{v'}$ on binding
energy (or detuning) indicate the strongly perturbed levels.
For the initial detuning of 4.1~cm$^{-1}$ (arrow labelled 0 in Fig.~\ref{fig:Brot}) and
a bandwidth of a picosecond pulse, perturbed levels are excited only for constant SOC.
In case of the $R$-dependent coupling of $W_1(R)$ and $W_2(R)$, regular
vibrational levels similar to those of the right-hand side of Fig.~\ref{fig:vibfcts}
are populated (see also Figs. 10 and 11 of Ref.~\cite{SlavaJCP99}).
In that case, it is expected that \textit{no}
deeply bound ground state molecules can be created
by applying the second (dump) pulse. In order to investigate whether 
formation of more deeply bound ground state molecules is possible at all
in case of $R$-dependent
SOC, the detuning of the pump pulse has been adjusted such that resonant levels
are excited (arrows labelled 1 and 2 in Fig.~\ref{fig:Brot}, cf. also Table~\ref{tab:SO}).
For larger detuning (7.6~cm$^{-1}$ and 9.2~cm$^{-1}$ as compared to 4.1~cm$^{-1}$),
the vibrational spacing becomes larger. A smaller number of vibrational levels
is then resonant within the bandwidth of the  PA pulse (for $\tau_P=10$~ps,
$\sim 10$ levels as compared to $\sim 20$ at $\Delta_P=4.1$~cm$^{-1}$). On the other hand,
the main reason for choosing a narrow-bandwidth pulse of $\tau_P=10$~ps was to
avoid excitation of the atomic resonance. This
becomes less likely for larger detuning even if
the bandwidth of the pulse is increased. Therefore, calculations for $\tau_P=10$~ps
are compared to $\tau_P=5$~ps in Table~\ref{tab:SO}. Both pulses have the same
pulse energy of 4.2~nJ. The larger number of vibrational levels resonant within
the bandwidth of the pulse for 5~ps compared to 10~ps
leads to a higher excitation probability. As in Sec.~\ref{sec:excitation}
(cf. Table~\ref{tab:pexc}),
the excitation probabilities reflect the dpms. The probability to populate
the last bound level of the \textit{ground} state decreases with increasing
detuning (cf. explanation in Sec.~\ref{subsec:efficiency}).

\begin{figure}[tb]
  \includegraphics[width=0.95\linewidth]{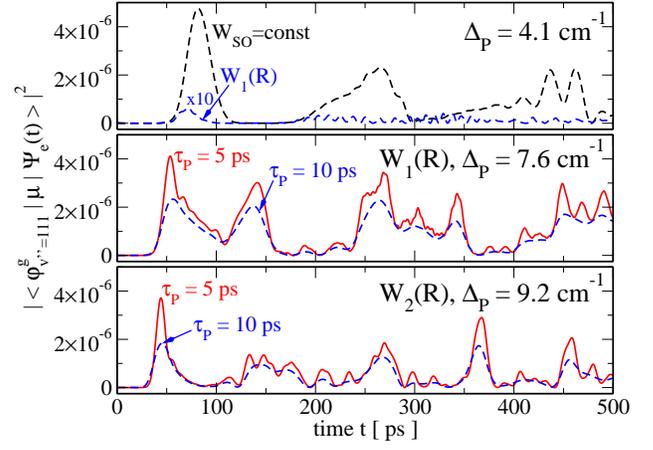}
  \caption{(Color online)
    Projection of the time-dependent excited state
    wavepacket onto the ground state level $v''=111$ for
    three different models of the spin-orbit coupling. The pump
    pulses are the same as in Table~\ref{tab:SO} with the
    detunings indicated in Fig.~\ref{fig:Brot}. In the
    calculations with $R$-dependent
    spin-orbit coupling, non-resonant excited state
    levels are populated
    in the upper panel while strongly perturbed
    levels are excited in the medium and bottom panel.
    \textit{Only after excitation of strongly perturbed levels,
    appreciable probability for transfer to more deeply bound ground state
    levels is obtained.}
  }
  \label{fig:proje2gSO}
\end{figure}

Fig.~\ref{fig:proje2gSO} shows the projection of the time-dependent excited
state wavepacket onto the ground state level $v''=111$,
$P_{v''=111}=|\langle \varphi^g_{v''=111} | \mu_\pi(\Op{R}) | \Psi_e(t)\rangle|^2$,
for the different SOC models and different pump detunings
($E_\mathrm{bind}^{v''=111}=5.74$~cm$^{-1}$, cf. Fig.~\ref{fig:ovlp} a and c).
The upper panel confirms that more deeply bound
ground state levels can only be populated
if resonant levels are excited in $0_u^+$: The
maximum value of $P_{v''=111}$ is 2 orders of magnitude smaller for
$W_1(R)$ than for $W_\mathrm{SO}=\mathrm{const}$. However, if the pump
detuning is adjusted such as to excite resonant levels, a similar
probability to populate more deeply bound levels is observed for
$W_1(R)$ and $W_2(R)$ as for  $W_\mathrm{SO}=\mathrm{const}$ (cf.
the maxima of  $P_{v''=111}$ in all three panels). Note that for larger
detuning, the best overlap of the excited state wavepacket is
obtained for $v''=110$ instead of  $v''=111$. The corresponding
binding energy is $E_\mathrm{bind}^{v''=110}=7.72$~cm$^{-1}$.
Therefore the resulting
dump detuning to populate this level,
$\Delta_D=\Delta_P-E_\mathrm{bind}^{v''=110}$, is very small
for $\Delta_P=7.6$~cm$^{-1}$ and $9.2$~cm$^{-1}$. In order to avoid
excitation of the atomic resonance by the dump pulse, $v''=111$ was
chosen as target level instead of $v''=110$. 

\begin{figure}[tb]
  \includegraphics[width=0.9\linewidth]{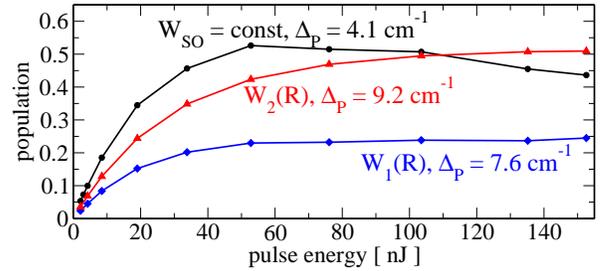}
  \caption{(Color online)
    Ground state population $P_g$ after 10~ps dump pulses
    resonant with $v''=111$ for three
    different models of the spin-orbit coupling $W_\mathrm{SO}$
    ($\tau_P=10$~ps for $\Delta_P=4.1$~cm$^{-1}$,
    $\tau_P=5$~ps for $\Delta_P=7.6$~cm$^{-1}$ and 9.2~cm$^{-1}$).
    \textit{The specific model of spin-orbit coupling  influences the exact amount
    of population which can be transferred to more deeply bound levels. 
    However, the scheme is robust and efficient over a large range of
    pulse energies for all  $W_\mathrm{SO}$.}
  }
   \label{fig:dump_SO}
\end{figure}

Fig.~\ref{fig:dump_SO} reports the results of time-dependent
calculations for the deexcitation step.
The pump-dump delay is chosen to correspond 
to the first maxima
in Fig.~\ref{fig:proje2gSO} ($W_\mathrm{SO}=\mathrm{const}$ with $\tau_P=10$~ps,
upper panel, $W_1(R)$ and $W_2(R)$ with $\tau_P=5$~ps, middle and lower
panel). The ground state
population after the second (dump) pulse is reduced
for $W_1(R)$ and $W_2(R)$ as
compared to constant SOC. However, the amount of
excited state population which can be transferred to $v''=111$
of the ground state, easily reaches 20\% for reasonable pulse energies
for all three coupling models. Furthermore, a saturation
of the transfer probability as a function of pulse energy is observed
for $R$-dependent SOC. This implies that the dump step is very robust
with respect to intensity fluctuations of the field.

\section{Conclusions}
\label{sec:concl}

Pump-dump photoassociation for rubidium below the $D_1$ line has been analyzed
with an emphasis on experimental feasibility. In particular, 
a setup such as in Refs.~\cite{Salzmann,Brown} was considered.
All potentials into which population can be excited by  a laser field were
included in the model. Both potentials and transition
dipole moments were based on \textit{ab initio} data and accurate long-range
expansions.

The first experiments on ultracold molecule formation with short laser
pulses~\cite{Salzmann,Brown} did not achieve the goal of creating molecules
from atoms in a MOT. Therefore, both pump and dump steps must each be optimized
in order to produce stable molecules in their singlet ground or lowest triplet state.
For the PA step, an efficient excitation mechanism is provided by the long-range
$1/R^3$-nature of several excited state potentials. However, 
the pump pulse is less efficient than one might initially expect
from experiments with CW lasers since a rather large
detuning from the atomic line is required for pulsed lasers. This is
due to the pulse bandwidth and the constraint
of not exciting the atomic resonance. \textit{The best compromise for pump pulses derived from
femtosecond oscillators is then found for pulse durations (FWHM) of 5~ps to 10~ps and
detunings of a few wavenumbers.}

The bandwidth $\Delta\omega_P$ of the pump pulse leads to the concept
of a photoassociation window which is comprised of the Franck-Condon points corresponding
to all resonant frequencies contained in $\Delta\omega_P$~\cite{ElianePRA04}.
The excitation is optimal if the photoassociation
window covers the range of internuclear distances of the last maximum of the last
bound level of the singlet ground / lowest triplet state. The Franck-Condon
radius corresponding to the central frequency (or detuning) of the pump pulse
$R_C(\Delta_P)$ should then be
close to the location $R_\mathrm{max}$ of the last maximum of
$\varphi^g_\mathrm{v=last}(R)$.
This is the case for pump pulse detunings of a few wavenumbers.
If one is interested only in exciting atoms,
pulses with positive frequency chirp perform best.
For subsequent formation of more deeply bound molecules
in their singlet ground or lowest triplet state,
a pump-dump scheme is required where
the optimal first pulse is transform-limited.
If the detuning from the atomic line is small enough,
the PA pulse transfers population also to  last bound levels of the
singlet ground or lowest triplet states, respectively.
 This has already been observed in the case of cesium~\cite{ElianePRA04}.
The creation of these extremely weakly bound molecules therefore
does not require a two-color pump-dump scheme. 

The use of two pulses can create
molecules in their singlet ground or lowest triplet state (bound by a few wavenumbers)
provided that an efficient stabilization mechanism exists. For the
Rb$_2$ states correlated to the $5S+5P_{1/2}$ asymptote,
such a mechanism was identified for $0_u^+$ leading to molecules in the
$X^1\Sigma_g^+$ ground state. 
A pump-dump scheme of photoassociation below the $D_1$
line via  the $0_u^+$ excited state provides then
an efficient means to create ground state
molecules bound by a few wavenumbers. In contrast, photoassociation via the $1_g$
and $0_g^-$ states will yield molecules in the lower triplet state which are extremely
weakly bound.
The excitation step from two atoms 
to the $0_u^+$ or  $1_g$ and $0_g^-$ excited states is unlikely to be selective.
% Population which is excited into the other two
% bound potentials correlated to the $5S+5P_{1/2}$ asymptote, $1_g$ and $0_g^-$,
% will be deexcited either into very weakly bound molecules in the lowest triplet state
% or into 'hot' scattering states corresponding to two untrapped atoms.
The efficient stabilization mechanism for  $0_u^+$ has been identified
as resonant spin-orbit coupling~\cite{ClaudePRL01,DulieuJOSA03}.
In a time-dependent process, resonant coupling
leads to a dynamical enhancement of stabilization making the dump step even more
efficient than deexcitation to Cs$_2$ lower triplet state molecules from  
$0^-_g(P_{3/2})$. Depending on the exact description of the spin-orbit coupling,
between 20\% to 50\% of the Rb$_2$ $0_u^+$ excited state
wavepacket can be
transferred to a single vibrational ground state level as compared to
14\% for  Cs$_2$ $0^-_g(P_{3/2})$~\cite{My05}.
In order to populate a single ground state level, dump pulses
should have a duration of 10~ps and be blue-detuned with respect to the
atomic resonance. The creation of a coherent superposition of ground state
levels is also possible.
This requires a broader bandwidth, i.e. shorter duration of the dump
pulses~\cite{My05}. 
However, care must then be taken to avoid excitation
of the atomic resonance.

The $0_u^+$ excited states contain both regular and strongly perturbed
levels. It was shown that the excitation of strongly perturbed levels
by the pump pulse is essential for deexcitation into more deeply bound
ground state levels. The position of the resonant levels in the vibrational
spectrum of $0_u^+$ and hence the required pump detuning
depends rather sensitively on the description of the
spin-orbit coupling.
Within our model, the position of these levels cannot be predicted
accurately.
However, the resonant levels are easily identified experimentally
by perturbations in the level spacings or rotational constants,
see e.g. Refs.~\cite{ManaaJCP02,BergemanPRA03}.
Accurate  spectroscopy of the $0_u^+$ states 
would improve both the potentials and the spin-orbit coupling and allow 
for obtaining a pump pulse from theory whose 
central frequency is at resonance with a strongly perturbed level.

The current work has been confined to study molecule formation via
excited states correlated to the $5S+5P_{1/2}$ asymptote. Some
conclusions can also be drawn with respect to the  $5S+5P_{3/2}$ asymptote, i.e.
the $D_2$ line. In that case, four attractive potentials into which
transitions can be induced by the laser field,
exist: $0_g^-$, $1_g$, $0_u^+$ and $1_u$. All four potentials
scale as $1/R^3$ at long range providing an efficient PA mechanism.
$0_g^-(P_{3/2})$ is known from PA with a CW laser~\cite{GabbaniniPRL00}
to provide an efficient stabilization mechanism due to the soft
repulsive wall of the long-range well. However, a technical difficulty
might prevent this route to be feasible: The depth of $0_g^-(P_{3/2})$ well
is only about 28~cm$^{-1}$, and pulse shaping over such a small
frequency range is hampered by the spectral resolution of the pulse
shaper.

An alternative route to ground state molecule formation might be
provided by employing a femtosecond frequency comb~\cite{Jun} which can
be operated in the picosecond to nanosecond regime. This would
avoid the problem of exciting the atomic resonance due to large pulse
bandwidth and allow for smaller detunings to be used. The pump step
could thus be significantly enhanced, recovering the efficiency of
PA with a CW laser.

In the present study, pump-dump photoassociation has been discussed
in the frame of a two-atom picture, with zero angular momentum and no transfer of
angular momentum from the light to the molecule.
One pair of pulses for excitation and stabilization was considered.
The calculation of absolute molecule formation rates is beyond the scope of
the present study. Such knowledge is important to
estimate the laser intensity required to produce a detectable
number of ground state molecules. In order to estimate absolute 
rates, two questions 
need to be addressed: (i) averaging over thermal
(translational and rotational) and angular distributions
of the $N$-atom system, and (ii) the problem of the repetition rate.
Answering the first question requires the solution of the pump-dump dynamics
for several initial scattering states (including higher collisional momentum
than $l=0$) and weighting the results with the
Boltzmann factors and the angles between the molecular and polarization
axes. The problem of the repetition rate stands for the fact that the
second pair of pump and dump pulses will act on a different initial state than
the first pair. In particular, the ground state population within the
resonance window might be decreased. It furthermore has to be ensured
that the second pair of pulses does not destroy the molecules created by the first
pair. In order to address these questions, the dynamics between the pulse
pairs has to be solved. These questions constitute the subject of ongoing work.

In view of prospective applications, vibrationally excited ultracold
molecules are not adequate. The next goal is therefore
to create ultracold molecules in their absolute rovibronic ground state.
Molecules at temperatures  below 1~mK are obtained by
assembling them from ultracold atoms due to interaction with an
external field. The molecule formation process is
coherent for magnetic~\cite{StreckerPRL03,XuPRL03,HerbigScience03} or
optical~\cite{MyPRL05} Feshbach resonances and for photoassociation with
short laser pulses.
Feshbach resonances and photoassociation with a single pulse lead to
molecules in one of the last bound ground state levels,
just below the dissociation limit. Such molecules typically have bond lengths
on the order of 100~$a_0$ and are bound by less than 0.1~cm$^{-1}$. At these
tiny binding energies, singlet and triplet character are mixed due to hyperfine
interaction. 
Two-color pump-dump photoassociation allows for creating
molecules which are more deeply bound. Bond lengths of 25~$a_0$ (15~$a_0$) and 
binding energies of
5-10~cm$^{-1}$ (110~cm$^{-1}$) were found for rubidium (cesium~\cite{My05}).
For subsequent Raman-type transitions to $v=0$,
molecules with bond lengths of 15-25~$a_0$ provide a much better
starting point than the very extended Feshbach molecules: Due to better
Franck-Condon overlaps, a smaller number of Raman steps and a reduced pulse energy
will be required~\cite{MyPRA04}.
The advantage of the $0_u^+$ route as shown in this study for rubidium
and of resonant spin-orbit coupling in general~\cite{SagePRL05} is that
molecules in their singlet, i.e. their absolute electronic ground state are
created. Therefore a two-color pump-dump scheme with picosecond pulses
provides an efficient first step toward
obtaining ultracold molecules in their absolute ground state.

\begin{acknowledgments}
  We are grateful to Olivier Dulieu for making
  the \textit{ab initio} data of the Rb$_2$ potentials and transition dipole moments
  available to us, to Eliane Luc-Koenig for encouragement and very
  fruitful discussions and to Roland Wester for his comments on the manuscript.
  C.P.K. would like to thank Sandy Ruhman for enlightening discussions on
  femtosecond laser technology 
  and acknowledges financial support
  from the Deutsche Forschungsgemeinschaft. This work has been supported by
  the European Commission in the frame of the 
  Cold Molecule Research Training Network under contract HPRN-CT-2002-00290.
  The Fritz Haber Center is supported
  by the Minerva Gesellschaft f\"{u}r die Forschung GmbH M\"{u}nchen, Germany.
\end{acknowledgments}

\end{document}